\documentclass{aa}  

\usepackage{amsmath}
\usepackage{amssymb}
\usepackage{graphicx}
\usepackage{color}
\usepackage{txfonts}
\usepackage{ulem}

\usepackage{comment}

\usepackage{xcolor}
\definecolor{xlinkcolor}{cmyk}{1,1,0,0}
 \usepackage[bookmarks=false,         
     pdfnewwindow=true,      
     colorlinks=true,    
     linkcolor=xlinkcolor,     
     citecolor=xlinkcolor,     
     filecolor=xlinkcolor,  
     urlcolor=xlinkcolor,      
final=true
 ]{hyperref}

\usepackage{savesym}
\savesymbol{tablenum}
\usepackage{siunitx}
\usepackage{xspace}
\restoresymbol{SIX}{tablenum}
\usepackage{hyperref}

\newcommand{\St}{\ensuremath{\mathrm{St}}\xspace}

\newcommand\dbquote[1]{\textquotedblleft #1\textquotedblright}
\newcommand\sgquote[1]{\textquotedblleft #1\textquotedblright}

\newcommand\partialdiff[1]{\frac{\partial}{\partial #1}}
\newcommand\stdiff[1]{\textrm{d} #1}

\newcommand\partiallogdiff[1]{\textrm{d}\mathrm{ln} #1/\textrm{d} \mathrm{ln}\, r}

\newcommand\Sigmagas{\Sigma_\textrm{g}}
\newcommand\Sigmadust{\Sigma_\textrm{d}}

\AtBeginDocument{}%
\AtBeginDocument{}%
\AtBeginDocument{}%
\AtBeginDocument{}%
\AtBeginDocument{}%

\begin{document} 
   \title{Millimeter emission in photoevaporating disks is determined by early substructures}
   \titlerunning{Dust evolution in photoevaporative disks}
   \author{Mat\'ias G\'arate\inst{1}
          \and
          Til Birnstiel\inst{2,3}
          \and
          Paola Pinilla \inst{4,1}
          \and
          Sean M. Andrews \inst{5}
          \and
          Raphael Franz\inst{2}
          \and
          Sebastian Markus Stammler\inst{2}
          \and
          Giovanni Picogna\inst{2}
          \and
          Barbara Ercolano\inst{2}
          \and
          Anna Miotello \inst{6}
          \and
          Nicol\'as T. Kurtovic \inst{1}
          }
   \institute{
   $^{1}$Max-Planck-Institut f\"ur Astronomie, K\"onigstuhl 17, 69117, Heidelberg, Germany\\
   $^{2}$University Observatory, Faculty of Physics, Ludwig-Maximilians-Universit\"at M\"unchen, Scheinerstr.\ 1, 81679 Munich, Germany\\
   $^{3}$Exzellenzcluster ORIGINS, Boltzmannstr. 2, D-85748 Garching, Germany\\
   $^{4}$Mullard Space Science Laboratory, University College London, Holmbury St Mary, Dorking, Surrey RH5 6NT, UK\\
   $^{5}$Center for Astrophysics | Harvard \& Smithsonian, 60 Garden Street, Cambridge, MA 02138, USA\\
   $^{6}$European Southern Observatory, Karl-Schwarzschild-Str. 2,85748 Garching, Germany\\
              \email{garate@mpia.de}
             }
   \date{}

  \abstract
   {
   Photoevaporation and dust-trapping are individually considered to be important mechanisms in the evolution and morphology of protoplanetary disks. However, it is not yet clear what kind of observational features are expected when both processes operate simultaneously.
   }
   {
   We studied how the presence (or absence) of early substructures, such as the gaps caused by planets, affects the evolution of the dust distribution and flux in the millimeter continuum of disks that are undergoing photoevaporative dispersal. We also tested if the predicted properties resemble those observed in the population of transition disks.
   }
   {
   We used the numerical code \texttt{Dustpy} to simulate disk evolution considering gas accretion, dust growth, dust-trapping at substructures, and mass loss due to X-ray and EUV (XEUV) photoevaporation and dust entrainment. Then, we compared how the dust mass and millimeter flux evolve for different disk models.
   }
   {
   We find that, during photoevaporative dispersal, disks with primordial substructures retain more dust and are brighter in the millimeter continuum than disks without early substructures, regardless of the photoevaporative cavity size.
   Once the photoevaporative cavity opens, the estimated fluxes for the disk models that are initially structured are comparable to those found in the bright transition disk population ($F_\textrm{mm} > 30\, \textrm{mJy}$), while the disk models that are initially smooth have fluxes comparable to the transition disks from the faint population ($F_\textrm{mm} < 30\, \textrm{mJy}$), suggesting a link between each model and population.
   }
   {
   Our models indicate that the efficiency of the dust trapping determines the millimeter flux of the disk, while the gas loss due to photoevaporation controls the formation and expansion of a cavity, decoupling the mechanisms responsible for each feature. 
   In consequence, even a planet with a mass comparable to Saturn could trap enough dust to reproduce the millimeter emission of a bright transition disk, while its cavity size is independently driven by photoevaporative dispersal.
   }

   \keywords{accretion, accretion disks -- 
            protoplanetary disks --
            hydrodynamics --  
            methods: numerical}

   \maketitle
%

\section{Introduction} \label{sec_Intro}

Observations of nearby star-forming regions reveal that the fraction of protoplanetary disks around young stellar objects decreases rapidly with age, indicating that the process of disk dispersal is relatively fast compared to the disk lifetime \citep[][]{Koepferl2013, Ribas2015}.
Theoretical models of disk evolution suggest that photoevaporation could explain this fast dispersal, which occurs when high energy photons 
in the far ultra-violet (FUV), extreme ultra-violet (EUV), and X-ray wavelength ranges of the spectra hit the gas particles on the disk surface and unbind them from the stellar gravitational potential.
When the mass loss rate due to photoevaporation exceeds the local accretion rate, a cavity opens and the disk enters into the photoevaporative dispersal regime, which can clear the remaining material on timescales of $\SI{e5}{yrs}$ from the inside out \citep[e.g.,][]{Clarke2001, Alexander2006a, Gorti2009a, Owen2010, Ercolano2017_Review}.\par

Photoevaporation has also been proposed as an explanation for transition disks, since the deficit of near-and mid-infrared (NIR and MIR) emission observed in the spectral energy distribution (SED) of these objects is linked to a lack of small grains in the inner regions \citep[][]{Strom1989, Skrutskie1990, Espaillat2014, vanderMarel2016}, which is consistent with a cavity in the inner disk such as those carved by photoevaporative dispersal \citep{Alexander2006b, Owen2011, Ercolano2018}, even though not all SED-selected disks might show a cavity in the millimeter continuum, or they might be related to highly inclined disks \citep[][]{vanderMarel2022}. 
Mixed models of photoevaporation and dead zones \citep[][]{Morishima2012, Garate2021} have also succeeded in explaining the high accretion rates found in transition disks \citep[e.g.,][]{Cieza2012, Alcala2014, Manara2014, Manara2016_a, Manara2016_b, Manara2017}, and the presence of a compact inner disk inside their cavities \citep[e.g.,][]{Kluska2018, Pinilla2019, Pinilla2021}, which was one of the main limitations of models of photoevaporation acting alone.\par

However, in addition to the infrared deficits, wide cavities, and high accretion rates, transition disks also seem to be distributed across two populations in terms of their millimeter flux, with a break at approximately $F_\textrm{mm} = \SI{30}{mJy} \cdot\, (d/\SI{140}{pc})^{-2}$ \citep[][]{Owen2012, Owen2016_review}. %
The fluxes of the population of faint transition disks can be easily reproduced by standard photoevaporation models with dust evolution \citep[][]{Owen2019, Garate2021}, but it is not yet clear if disks undergoing photoevaporation could also reproduce the fluxes of the millimeter bright population. \par

Observations of rings and gaps in protoplanetary disks \citep[e.g.,][]{ALMA2015, Andrews2018, long2018, cieza2021} indicate that disk density profiles are rich in substructures that can act as dust traps \citep{Whipple1972, Weidenschilling1977}, and they greatly affect the resulting dust distribution and flux in the millimeter continuum \citep[e.g.,][]{Pinilla2012, Pinilla2012_b, Pinilla2020}. 
However, models that include consistent dust evolution \citep[i.e., growth, fragmentation, and multiple species, e.g.,][]{Birnstiel2010, Birnstiel2012, Drazkowska2019}, early substructures\footnote{In this article, we refer to substructures that are present in the disk before the onset of photoevaporative dispersal, as \dbquote{early} or \dbquote{primordial}.}, along with photoevaporative dispersal have not been widely studied. 
To our knowledge, only the work of \citet[][]{Booth2020} has simultaneously considered all three of the mentioned ingredients, in the specific context of the Solar System formation, where the authors show that large amounts of dust can be trapped by Jupiter, therefore decreasing the amount of refractory material delivered to the Sun by the time that photoevaporative dispersal starts to clear out the disk.
Thus, it is necessary to further determine the emission in the millimeter continuum of photoevaporating disks where early dust traps are included, and compare them with the fluxes and morphology found in the bright transition disk population.\par

We note that a common issue of photoevaporative disk models is that these tend to overpredict the fraction of non-accreting transition disks, colloquially dubbed as \dbquote{relic disks} \citep[][]{Owen2011}, which have not yet been detected by observations. In fact, based on the current observational thresholds, the fraction of relic disks should only be around $3\%$ of the total transition disk population \citep[][]{Hardy2015}, though this fraction should be revised, since several of the observed systems have been identified as non-cluster members in recent years \citep[see][]{Michel2021}.
Mechanisms that remove dust grains from the disk, such as radiation pressure \citep{Owen2019} or wind entrainment \citep[][]{Franz2020}, could in principle reduce the infrared signal of these relics below detection limits, though it is an open question whether or not these processes would be able to remove enough solid material in a disk where early dust traps were present. Alternatively, it could also be that the fraction of relic disks is simply lower than previously predicted \citep[see][]{Ercolano2018, Garate2021}.
\par

In this paper, we studied the evolution of a photoevaporating disk from the point of view of the dust dynamics, using a 1D model. In particular, we focused on how the presence of early substructures (such as the ones caused by planets) affects the resulting dust density and size distribution during photoevaporative dispersal, along with the predicted flux in the millimeter continuum ($\lambda = \SI{1.3}{mm}$), and SED in the infrared.
In our model we included the growth and fragmentation of multiple dust species \citep[][]{Birnstiel2010}, state-of-the-art models of X-ray and EUV photoevaporation \citep[][]{Picogna2019}, and the loss of dust particles with the photoevaporative winds \citep[][]{Franz2020}. \par

We further discuss whether the predicted observational signatures of photoevaporating disks can be linked to those observed in transition disks, in terms of millimeter flux and morphology, how the presence or absence of substructures and their properties (location and amplitude) affects the observable features of dispersing disks; and finally, if the dust loss by wind entrainment during the dispersal process can explain why relic disks have not yet been detected.\par

In \autoref{sec_Theory} we introduce our disk evolution model and its implementation.
In \autoref{sec_Setup} we present our simulation setup, and the explored parameter space.
Our results are shown in \autoref{sec_Results}, and in \autoref{sec_Discussion} we discuss them in the context of observations. 
We summarize our results in \autoref{sec_Summary}.
%

\section{Disk model} \label{sec_Theory}
In this section we present our disk evolution model in 1D, which includes gas and dust advection, dust diffusion, X-ray photoevaporation, a prescription for a gap-like substructure, and the evolution of the dust size distribution through coagulation and fragmentation; all  assuming that the disk is axisymmetric.

\subsection{Gas evolution} \label{sec_GasEvolution}
The gas evolution is governed by the viscous diffusion and the mass loss due to photoevaporation. Then, the evolution of the gas surface density $\Sigmagas$ can be described through the following diffusion equation \citep{Lust1952, Pringle1981}
\begin{equation} \label{eq_GasEvolution}
    \partialdiff{t}\Sigmagas = \frac{3}{r}\partialdiff{r}\left(r^{1/2} \partialdiff{r}\left(\nu \Sigmagas r^{1/2} \right) \right) - \dot{\Sigma}_w,
\end{equation}
where $r$ is the radial distance to the central star and $\dot{\Sigma}_w$ is the mass loss rate, which depends on the X-ray luminosity $L_x$ from the central star (see \autoref{sec_PhotoModel}).
The gas viscous evolution is characterized by the kinematic viscosity $\nu$, which is defined in \citet{Pringle1981} as
\begin{equation} \label{eq_ViscNu}
    \nu = \alpha c_s^2 \Omega_k^{-1},
\end{equation}
where $\alpha$ is a dimensionless parameter that represents the magnitude of the gas turbulence \citep{Shakura1973}, $c_s = \sqrt{k_B T / \mu m_p}$, is the isothermal sound speed, with $m_p$ the proton mass, $\mu = 2.3$ the mean molecular weight, $k_B$ the Boltzmann constant, $T$ the gas temperature. $\Omega_k = \sqrt{G M_*/r^3}$, is the Keplerian orbital speed, with $G$ the gravitational constant, and $M_*$ the mass of the central star.\par

To induce a gap-like substructure in the gas surface density, such as the one that would be created by a planet, we implemented the following radial turbulence profile that includes a Gaussian bump
\begin{equation} \label{eq_alpha_bump}
    \alpha(r) = \alpha_0 \times \left(1 + A_\textrm{gap} \exp\left(-\frac{\left(r - r_\textrm{gap}\right)^2}{2w_\textrm{gap}^2}   \right)\right),
\end{equation}
where $\alpha_0$ is the base value for the turbulence, $r_\textrm{gap}$ is the location of the gap structure, $w_\textrm{gap}$ is the Gaussian standard deviation that controls the gap width, and $A_\textrm{gap}$ is the amplitude of the bump, which in turn controls the depth of the gap in the gas surface density profile. 
This Gaussian factor was also used by \citet{Pinilla2020}, though in their study it was used to create bumps in the surface density, instead of gaps.\par

Physically, a local increment in the kinematic viscosity creates a region where the gas diffuses faster, which translates into a gap in the surface density profile. For disks that are in steady state, the gas accretion rate is radially constant, and given by $\dot{M}_\textrm{g} = 3 \pi \Sigmagas \nu$ \citep{Pringle1981}. The relation between the gas surface density and viscosity can also be applied to disks that are in quasi-steady state \citep[such as the self-similar solution described by][]{Lynden-Bell1974}, where variations in the $\alpha$ turbulence profile yield inversely proportional variations in the gas surface density, approximately following $\Sigmagas \propto \alpha^{-1}$ \citep[see examples in][among others]{Dullemond2018, Stammler2019, Pinilla2020}.\par

For the gas temperature we assumed that the disk is heated passively by the central star, and therefore its temperature profile is related to the stellar temperature $T_*$ and size $R_*$ through
\begin{equation} \label{eq_Temperature}
    T = \theta_{irr}^{1/4} \left(\frac{r}{R_*}\right)^{-1/2} T_*,
\end{equation}
where $\theta_{irr} = 0.05$ is the disk irradiation angle.

\subsection{Dust evolution}
For the dust evolution model, we followed the work of \citet{Birnstiel2010}, which describes the advection, diffusion, coagulation, and fragmentation of multiple dust species, in response to the interaction with the gas component through the aerodynamic drag force.\par

The corresponding advection-diffusion equation for the dust surface density $\Sigmadust$ is
\begin{equation} \label{eq_dust_advection}
    \partialdiff{t} \left(r \, \Sigmadust \right) + \partialdiff{r} (r \, \Sigmadust \, v_{\textrm{d}}) - \partialdiff{r} \left(r D_\textrm{d} \Sigmagas \partialdiff{r}\epsilon \right) = - \dot{\Sigma}_\textrm{w,d},
\end{equation}
where $v_\textrm{d}$ corresponds to the dust radial velocity, $D_\textrm{d}$ is the dust diffusivity, $\epsilon = \Sigmadust/\Sigmagas$, is the local dust-to-gas ratio, and $\dot{\Sigma}_\textrm{w,d}$ is the dust loss rate due to wind entrainment \citep{Hutchison2016, Franz2020, Hutchison2021, Booth2021, Franz2022a, Franz2022b}. 
Here we note that \autoref{eq_dust_advection} acts on every individual dust species, and that all dust related quantities are defined as functions of the particle size $a$.\par

All components of the dust dynamics for a given particle size are determined by their dimensionless stopping time, the Stokes number
\begin{equation} \label{eq_StokesMidplane}
    \textrm{St} = \frac{\pi}{2}\frac{a\, \rho_s}{\Sigma_g} \cdot
    \begin{cases}
				1 & \lambda_\textrm{mfp}/a \geq 4/9\\
                \frac{4}{9} \frac{a}{\lambda_\textrm{mfp}} & \lambda_\textrm{mfp}/a < 4/9.
	\end{cases} 
\end{equation}
This definition distinguishes between the Epstein drag regime, that dominates when the grain sizes are smaller than the mean free path $\lambda_\textrm{mfp}$, and the Stokes regime, that occurs when the particles are large or the gas is very dense (for example in the inner disk).\par

The radial velocity of a dust particle is then
\begin{equation} \label{eq_dust_radial_velocity}
    v_\textrm{d} = \frac{1}{1 + \mathrm{St}^2} v_\nu -  \frac{2 \mathrm{St}}{1 + \mathrm{St}^2} \eta v_k,
\end{equation}
following \citet{Nakagawa1986} and \citet{Takeuchi2002}, where $v_\nu$ is the gas advection velocity due to viscous diffusion and $\eta = -\, (1/2)\,  (h_\textrm{g} / r)^2\, \partiallogdiff{P}$, is the relative difference between the Keplerian velocity $v_k$ and the gas orbital velocity, due to its own pressure support. The isothermal pressure is defined as $P = \rho_\textrm{g,0} c_s^2$, with $\rho_\textrm{g,0}$ the gas volume density at the midplane, and $h_\textrm{g} = c_s \Omega_k^{-1}$, is the gas scale height.\par

Finally, the dust diffuses with a diffusivity $D_\textrm{d} = \nu / (1 + \textrm{St}^2)$ \citep{Youdin2007}, where we note that $D_\textrm{d} \approx \nu$ for particles with $\St \ll 1$. \par

In addition to advection and diffusion, dust species also grow and/or fragment depending on their relative velocities and their collision rate, where the evolution of the grain size distribution is governed by the Smoluchowski equation \citep{Birnstiel2010}. In a typical protoplanetary disk there are two regimes of dust growth: fragmentation limited, or drift limited \citep{Birnstiel2010, Birnstiel2012}. \par

The fragmentation limit occurs when the collision velocities of larger dust grains surpass the fragmentation velocity threshold $v_\textrm{frag}$ (which depends on the material properties), resulting in destructive collisions, and replenishing the population of small grains \citep[e.g.,][]{Ormel2007, Brauer2008, Birnstiel2009}.
This regime can be typically found in the inner regions of protoplanetary disks, regions with high turbulence, and in pressure maxima, where the drift velocity is zero, and is given by
\begin{equation} \label{eq_fragmentation_limit}
    \mathrm{St}_{\textrm{frag}} = \frac{1}{3}\frac{v_\textrm{frag}^2}{\alpha c_s^2}.
\end{equation}\par

The drift limit, on the other hand, occurs when the dust grains grow to the point where the drift timescale is shorter than the growth timescale \citep{Birnstiel2010}. This regime appears in regions with steep pressure gradients, such as the outer disk and regions with low dust-to-gas ratios, where the maximum size that a grain can grow to is approximately
\begin{equation} \label{eq_drift_limit}
    \mathrm{St}_{\textrm{drift}} = \left|\frac{\textrm{dln}\, P}{\textrm{dln}\, r }\right|^{-1} \frac{v_k^2}{c_s^2} \epsilon_\textrm{tot},
\end{equation}
where $\epsilon_\textrm{tot}$ refers to the \textit{local} dust-to-gas ratio of all dust species combined.
\subsection{Photoevaporation model} \label{sec_PhotoModel}
\begin{figure}
\centering
\includegraphics[width=90mm]{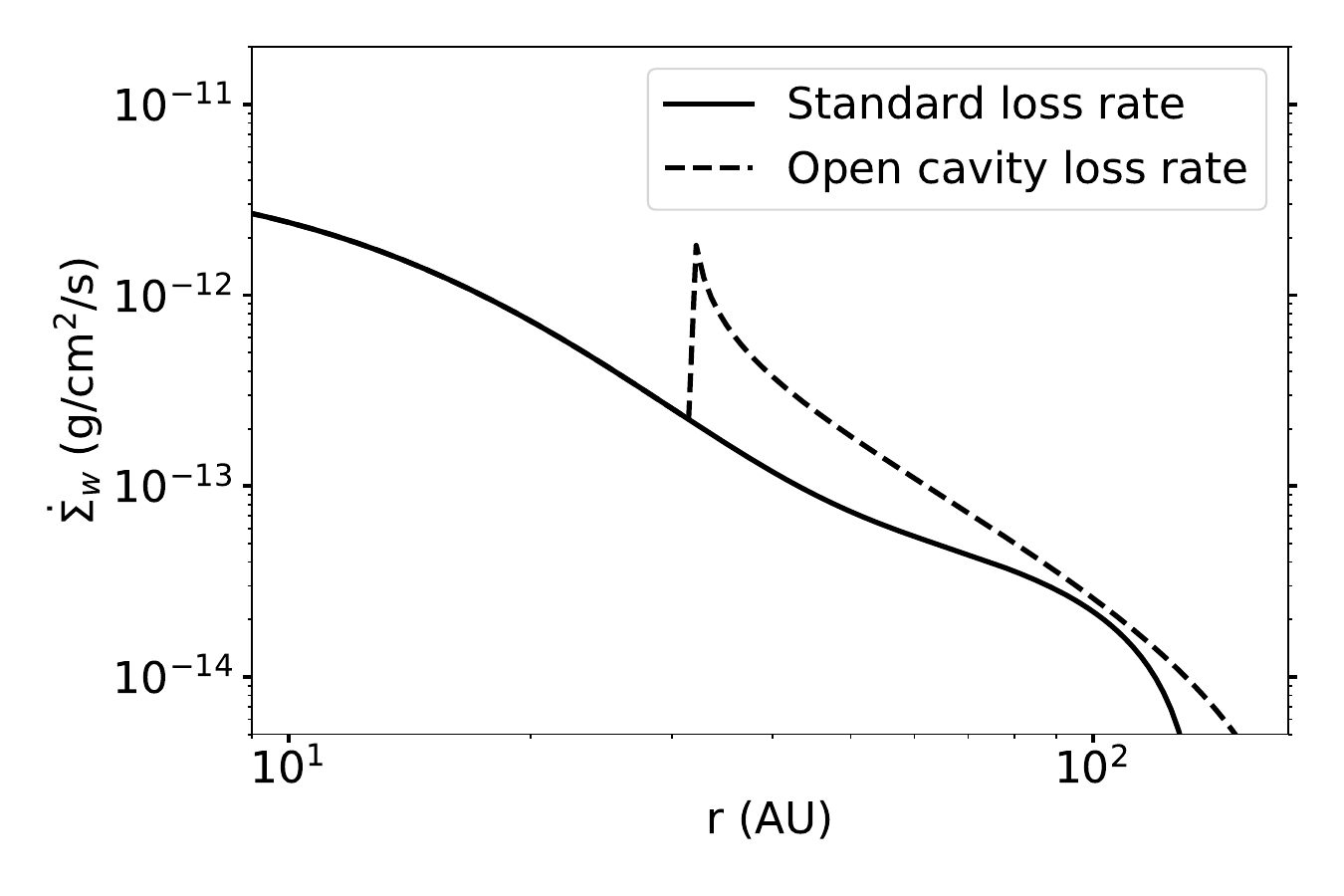}
 \caption{
 Example of the gas loss rate profile, following the photoevaporation model from \citet[][their Equations 2 and 4]{Picogna2019}, for $L_x = \SI{e30}{erg\, s^{-1}}$. The figure shows the mass loss rate before the photoevaporative cavity opens (solid line) and after the photoevaporative cavity opens (dashed line), with the cavity edge located at $r_\textrm{cavity} \approx \SI{30}{AU}$ after \SI{1.9}{Myr} of evolution.
 }
 \label{Fig_LossRateProfile}
\end{figure}

When high energy radiation from the central star hits the disk surface layer, the material is heated and unbound from the stellar gravitational potential in a process called photoevaporation, which ultimately leads to disk dispersal \citep{Clarke2001, Alexander2006a, Alexander2006b, Alexander2007}.\par

In our model, we implemented the mass loss rate profile from \citet[][see their Equations 2-5]{Picogna2019} into the sink term $\dot{\Sigma}_w$ of the gas diffusion equation (\autoref{eq_GasEvolution}) to simulate the effect of X-ray and EUV photoevaporation, where the total mass loss rate increases with the stellar X-ray luminosity $L_x$.\par

\autoref{Fig_LossRateProfile} shows an example of the $\dot{\Sigma}_w (r)$ profile, that distinguishes between the case when the disk is still young and without a cavity, and the case after the photoevaporative cavity opens, where the cavity edge is directly irradiated by the central star.
This model is valid for a $\SI{0.7}{M_\odot}$ star irradiating a disk with the X-ray spectrum as given by \citet{Ercolano2009}. \citet{Ercolano2021} and \citet{Picogna2021} expand this model to a range of stellar masses and apply observationally determined X-ray spectra. Given that the new models are qualitatively similar to those of \citet{Picogna2019} used here, we do not expect that their implementation in our work would lead to significant changes in our conclusions.\par

From \autoref{eq_GasEvolution} we see that the gas evolution can be dominated by either viscous diffusion or by photoevaporation. In the early stages, when the disks are more massive, the viscous accretion is generally thought to be the dominant evolution mechanism, while photoevaporation will dominate the disk dispersal in later stages, removing the remaining material by opening a cavity from the inside out \citep[][]{Clarke2001}.\par

Along with the gas removal through photoevaporative winds, we can also expect for a fraction of the dust grains to be entrained in the photoevaporative flow \citep[][]{Hutchison2016, Franz2020}. To quantify the dust loss rate we define a sink term 
in \autoref{eq_dust_advection}
\begin{equation} \label{eq_dust_entrainment}
   \dot{\Sigma}_\textrm{w,d} = \epsilon_\textrm{w}  \dot{\Sigma}_\textrm{w},
\end{equation}
where $\epsilon_\textrm{w}$ represents the dust-to-gas \textit{loss} ratio, and 
is defined in our model as the mass fraction of particles that are small enough to couple to the gas motion with $a \leq a_\textrm{w}$, and that lie above the wind launching surface with $z \geq h_\textrm{w}$, where $h_\textrm{w}$ and $a_\textrm{w}$ are free parameters in our model.\par

The final ingredient to find $\epsilon_\textrm{w}$ is to define a vertical structure for the gas and dust volume densities $\rho_\textrm{g}$ and $\rho_\textrm{d}$. 
Following \citet{Fromang2009}, and assuming that the gas is in vertical hydrostatic equilibrium (with constant temperature in the vertical direction), we model the vertical density distribution as
\begin{equation} \label{eq_vertical_density}
\rho_\textrm{g,d}(z) = \frac{\Sigma_\textrm{g,d}}{\sqrt{2\pi} h_\textrm{g,d}} \exp\left(-\frac{z^2}{2 h^2_\textrm{g,d}}\right),
\end{equation}
where the dust scale height $h_\textrm{d}$ is
\begin{equation}\label{eq_scaleheight_dust}
h_\textrm{d} = h_\textrm{g} \cdot \min\left(1, \sqrt{\frac{\alpha}{\min(\mathrm{St},1/2) (1+\mathrm{St}^2)  }}\right),
\end{equation}
following \citet{Youdin2007} and \citet{Birnstiel2010}, where we notice that the dust scale height is smaller than the gas one, since large grains with ($\textrm{St} \gtrsim \alpha$) tend to settle toward the midplane.\par

The formal expression for the dust-to-gas loss ratio can then be written as
\begin{equation}\label{eq_lossratio}
    \epsilon_\textrm{w} (a) = \frac{\int_{h_\textrm{w}}^{\infty} \rho_\textrm{d}(z, a) \stdiff{z}}{\int_{h_\textrm{w}}^{\infty} \rho_\textrm{g}(z) \stdiff{z}},
\end{equation}
which due to settling and growth is always smaller than the local dust-to-gas mass ratio ($\epsilon_\textrm{w} < \epsilon$).\par

Our photoevaporation model assumes that all the small grains above the wind launching surface are fully entrained and that none of the removed material (gas or dust) falls back onto the disk \citep[][]{Clarke2016, Picogna2019, Franz2020, Sellek2021}.

\section{Simulation setup} \label{sec_Setup}
We performed our simulations using the code \texttt{DustPy}\footnote{
A legacy version of the code was used.
The latest version of \texttt{DustPy} is available on \href{https://github.com/stammler/DustPy}{github.com/stammler/DustPy}. 
}$^{,}$\footnote{
\texttt{DustPy} is built on the \texttt{Simframe} simulation framework \citep[\href{https://github.com/stammler/simframe}{github.com/stammler/simframe},][]{Stammler2022_Simframe}.}
\citep[][]{Stammler2022_Dustpy}, that solves the gas and dust surface density evolution (\autoref{eq_GasEvolution} and \autoref{eq_dust_advection}) following the  model from \citet{Birnstiel2010}.\par

We implemented the photoevaporation model described in \autoref{sec_PhotoModel} and induced gap-like substructures using the $\alpha$ turbulence profile from \autoref{eq_alpha_bump}. 
For our study, we focused on the impact that the X-ray luminosity $L_x$, the gap amplitude $A_\textrm{gap}$, and the gap location $r_\textrm{gap}$ have on the disk evolution.\par

For each simulation we tracked the evolution of the disk mass $M_\textrm{g,d}$ (in gas and dust), photoevaporative cavity size $r_\textrm{cavity}$, flux in the millimeter continuum $F_\textrm{mm}$ at $\SI{1.3}{mm}$, and SED.  
Our goal is then to compare these properties to the values observed in protoplanetary disk populations. In particular, we want to determine if these photoevaporating disks could be bright enough to explain the transition disk millimeter fluxes, or if they are too faint to be detected at all, and therefore related to the relic disk problem \citep{Owen2011, Owen2012, Owen2019}.\par

In this section we describe the initial conditions and numerical grid setup, the radiative transfer model that we used to post-process the simulations and obtain the millimeter fluxes and SEDs, and finally our parameter space exploration, which we used to study the impact of the X-ray luminosity and gap properties.
\subsection{Initial conditions and numerical grid} \label{sec_Setup_InitialCondition}
In our simulations, the central star has a mass of $M_* = \SI{0.7}{M_\odot}$, a radius of $\SI{1.7}{R_\odot}$, and a temperature of $\SI{4500}{K}$, selected to match the stellar parameters from the photoevaporative models of \citet{Picogna2019} and \citet{Owen2019}. For these stellar properties, the disk has a temperature of $\approx \SI{190}{K}$ at $\SI{1}{AU}$ (see \autoref{eq_Temperature}).\par

For the initial gas surface density profile we used a modified version of the \citet{Lynden-Bell1974} self-similar solution
\begin{equation} \label{eq_LBPprofile}
    \Sigmagas(r) = \frac{M_\textrm{disk}}{2\pi r_c^2} \left(\frac{r}{r_c}\right)^{-1} \exp(-r/r_c) \frac{\alpha_0}{\alpha(r)},
\end{equation}
where $M_\textrm{disk} = \SI{0.05}{M_\odot}$ is the total disk mass, and $r_c = \SI{60}{AU}$ is the disk characteristic radius.\par

To introduce a gap in the surface density profile from the beginning of the simulation, we added the factor  $\alpha_0/\alpha(r)$ to the self-similar solution. Then, the resulting gap structure is consistent and sustained by the turbulence profile defined in \autoref{eq_alpha_bump} \citep[][]{Stammler2019, Pinilla2020, Stadler2022}, where the gap center is located at $r_\textrm{gap}$ and the amplitude (i.e., the depth of the perturbation in the surface density profile) is determined by $A_\textrm{gap}$.
In the case where the gap amplitude is $A_\textrm{gap}=0$ (i.e., no gap), we recover the traditional self-similar solution. We used a value of $\alpha_0 = \SI{e-3}{}$ for the disk base turbulence.\par

The initial dust-to-gas ratio is $\epsilon_0 = \SI{1.5e-2}{}$, and the initial dust size distribution follows the MRN distribution \citep{Mathis1977}, with an initial maximum grain size of $a_0 = \SI{1}{\mu m}$. 
Our initial dust-to-gas ratio is higher than the canonical $1\%$, motivated by a recent study by \citet{Lebreuilly2020}, which indicates that protoplanetary disks may heritage higher dust-to-gas ratio than the ISM from the protostellar collapse.\par

For the dust grains we assumed that these are compact and covered by ice, with a material density of $\rho_s = \SI{1.6}{g\, cm^{-3}}$ and a fragmentation velocity of $v_\textrm{frag} = \SI{10}{m\, s^{-1}}$ \citep{Wada2011, Gundlach2011, Gundlach2015}, though notice that recent results suggest that the fragmentation velocity of icy grains could be lower than previously thought \citep{Gundlach2018, Musiolik2019, Steinpilz2019}.\par

We used a logarithmically spaced radial grid going from $\SI{4}{AU}$ to $\SI{300}{AU}$ with $n_r = 200$ radial cells, and a logarithmically spaced mass grid from $\SI{e-12}{g}$ to $\SI{e5}{g}$ (approx. \SI{0.5}{\mu m} to \SI{20}{cm} in grain sizes) with $n_m = 120$ cells.
Finally, in order to determine the fraction of dust that is entrained in the photoevaporative wind $\epsilon_\textrm{w}$, we employed a 1+1D approach in which we constructed a vertical grid locally at every radial grid cell to solve the integrals in \autoref{eq_lossratio}. This grid is defined in function of the gas scale height, going from the midplane to $10\, h_\textrm{g}$ with $n_z = 100$ cells.\par

We saved the simulation outputs every \SI{0.1}{Myr}, and terminated the simulations when the photoevaporative cavity exceeded \SI{120}{AU} in size, since other photoevaporation regimes are more likely to become dominant over the X-ray driven dispersal for larger cavity sizes \citep[e.g. FUV,][]{Gorti2009a}. 
\subsection{Radiative transfer and optically thin approximation}
To obtain the millimeter fluxes $F_\textrm{mm}$ at $\lambda = \SI{1.3}{mm}$ we can take two approaches: the vertical slab approximation, or the complete radiative transfer calculation.
For the vertical slab approximation we used the vertically integrated surface density to calculate the optical depth $\tau_\nu = \sum_a \kappa_\nu(a) \Sigmadust(a)$, where $\kappa_\nu (a)$ is the absorption opacity and $\nu$ is the frequency, and obtain the total flux at $\lambda = \SI{1.3}{mm}$ ($\nu = \SI{230}{GHz}$) with
\begin{equation} \label{eq_Flux_OpticallyThin}
    F_\textrm{mm} = \int B_\nu(T) \left(1 - \exp(-\tau_\nu) \right) \stdiff{\Omega}, 
\end{equation}
where $B_\nu$ is the Planck function, $\stdiff{\Omega}$ is the solid angle differential, and $T$ is the vertically isothermal dust temperature (\autoref{eq_Temperature}).
This approach is ideal to quickly compute the fluxes for all snapshots directly from the dust distribution, but has the drawback that it is only reliable for low optical depths, it neglects the effect of self-scattering, and that the temperature profile may not be consistent with that of a irradiated disk, especially at the edge of the photoevaporative cavity.\par 

To obtain more accurate fluxes for key snapshots, we used the radiative transfer code \texttt{RADMC-3D}\footnote{\href{https://www.ita.uni-heidelberg.de/~dullemond/software/radmc-3d/}{www.ita.uni-heidelberg.de/~dullemond/software/radmc-3d/}} \citep{RADMC3D2012}, to recalculate the dust temperature, the millimeter fluxes at $\SI{1.3}{mm}$, and the SED between $\SI{0.1}{\mu m}$ and $\SI{1}{cm}$. For our calculations we considered the complete treatment of scattering, that includes polarization and anisotropy \citep{Kataoka2015}.
The radiative transfer is performed on a azimuthally symmetrical spherical grid, where the radial coordinate matches the logarithmically spaced grid from \texttt{Dustpy} (see \autoref{sec_Setup_InitialCondition}), and the colatitude coordinate covers the entire domain with $n_\theta = 180$ grid cells.
We used $\SI{e7}{}$ photon packages to calculate the thermal structure, $\SI{2.5e5}{}$ photon packages to calculate the emission in the millimeter continuum, and $\SI{e4}{}$ photon packages to calculate the SEDs. To account for the full scattering treatment we subdivided the azimuthal coordinate in $n_\phi=64$ grid cells.
For this work we also assumed that our disks are face-on (the inclination is $i = 0$), and that they are located at a distance of $d = \SI{140}{pc}$, which is the typical distance of the nearby star-forming regions \citep[][]{Dzib2018_Gaia2, Roccatagliata2020_Gaia2}.\par

For both the optically thin and the radiative transfer setup we used the opacity model from the DSHARP survey \citep{Birnstiel2018}, which assumes compact grains composed of water ice, troilite, refractory organics, and astronomical silicates \citep{henning1996, Draine2003, warren&brandt08}.
Then we used the code \texttt{OpTool}\footnote{  \href{https://github.com/cdominik/optool}{github.com/cdominik/optool}}\citep{Dominik2021} to obtain the opacities, following the Mie theory for compact grains, for all 120 grain sizes tracked by the \texttt{Dustpy} simulations.\par

While the DSHARP opacities provide a convenient framework that is common to several recent studies, it is not clear whether these represent the true absorption of dust grains accurately. For example, the model from \citet{Ricci2010}, based on the optical constants of \citet{Zubko1996}, \citet{ Draine2003}, and \citet{warren&brandt08}, leads to absorption opacities that are approximately one order of magnitude higher than the DSHARP opacities in the millimeter continuum, which leads to higher optical depths and fluxes \citep[][]{Zormpas2022, Stadler2022}.
Since the fluxes obtained from radiative transfer calculations are dependent on the selected opacity model, we included in our results a comparison between the fluxes obtained with the \citet{Birnstiel2018} and the \citet{Ricci2010} opacity models.
\subsection{Parameter space: X-ray luminosity and gap properties}
\begin{table}
 \caption{Parameter space.}
 \label{Table_Param}
 \centering
  \begin{tabular}{ l  c }
    \hline \hline
    \noalign{\smallskip}
    Variable & Value  \\
    \hline
    \noalign{\smallskip}
    $L_x$ [$10^{30}$ erg s$^{-1}$] & 0.3, \textbf{1.0}, 3.0  \\
    $r_\textrm{gap}$ [AU] & 20, \textbf{40}, 60   \\
    $A_\textrm{gap}$ & \textbf{0}, 1, 2, \textbf{4} \\
    \hline
  \end{tabular}
\end{table}

In this study we want to understand what effect the presence or absence of substructure has on the disk observable quantities during photoevaporative dispersal.
To explore the parameter space we selected two fiducial simulations: one without a gap ($A_\textrm{gap} = 0$), and one with a gap ($A_\textrm{gap} = 4$, i.e., a decrease by a factor of $0.2$ in the local $\Sigmagas$) located at $r_\textrm{gap} = \SI{40}{AU}$, where both simulations have the X-ray luminosity $L_x = \SI{e30}{erg\, s^{-1}}$.
For reference, a gap located at $\SI{40}{AU}$ and with an amplitude of $4$, is what we would expect from a planet of $\SI{225}{M_\oplus}$ (approx. twice the mass of Saturn), or planet-to-star mass ratio  of $q \approx \SI{9.5e-4}{}$, following the \citet[][]{Kanagawa2017} gap model.\par

Afterward, we repeated our study for different X-ray luminosities, while keeping the fiducial gap properties. Finally we studied the effect of the different gap locations and amplitudes, this time keeping the fiducial X-ray luminosity.
\autoref{Table_Param} shows the X-ray luminosity and gap properties of our parameter space, with the fiducial values in boldface. \par

The X-ray luminosities were selected from within the range of the Taurus luminosity distribution \citep{Preibisch2005}. Each value of $L_x$ can also be understood in terms of the resulting total mass loss rates $\dot{M}_w$, which are respectively $\SI{4.6e-9}{}$, $\SI{1.6e-8}{}$ and $\SI{3.2e-8}{M_\odot\, yr^{-1}}$ for the values listed in \autoref{Table_Param} \citep[see Eq. 5 from][]{Picogna2019}.\par

The gap locations were selected to be within (or at) the disk characteristic size $r_c = \SI{60}{AU}$, and the maximum gap amplitude was selected to ensure that dust trapping is effective at the different gap locations.
For the gap widths, we chose to apply a simple prescription of
\begin{equation}\label{eq_gap_widths}
    w_\textrm{gap} = 5 \left(\frac{r_\textrm{gap}}{\SI{40}{AU}}\right) \SI{}{AU},
\end{equation}
which roughly matches the widths of the dust traps from \citet{Pinilla2020}, and is always larger than the local scale height for our parameter space.\par

Finally, the amount of dust removed by photoevaporation in our model depends both on the maximum entrainment size and on the scale height of the wind launching region.
As fiducial values for our simulations, we assumed that only particles smaller than $a \leq a_\textrm{w} = \SI{10}{\mu m}$ \citep{Hutchison2016, Franz2020, Hutchison2021, Booth2021} can be carried by the photoevaporative winds, and that the particles must be at least above $z \geq h_\textrm{w} = 3 h_\textrm{g}$, though the photoevaporative surface can be located at higher altitudes.\par

In \autoref{Sec_Appendix_DustLoss} we further explored the parameter space for $a_\textrm{w}$ and $h_\textrm{w}$, though we do not expect for the resulting dust distribution to be greatly affected by the exact parameter values, since due to grain growth and settling dust loss should be small in comparison to the gas loss \citep[i.e., $\epsilon_\textrm{w} < \epsilon$, $ $][]{Franz2022a}. 
Motivated by the possibility of more efficient dust removal due to additional mechanisms such as radiation pressure \citep{Owen2019}, an FUV component in the wind \citep[e.g.,][]{Gorti2009a}, or grains lifted by magneto-hydrodyamical (MHD) winds \citep{Miyake2016}, we also included a model in which all the dust is fully entrained in the wind ($\epsilon_\textrm{w} = \epsilon$). We also included a comparison of the dust mass evolution of our fiducial model against the entrainment prescription of \citet{Booth2021}.

\section{Results} \label{sec_Results}
%
\subsection{Fiducial models}
\begin{figure}
\centering
\includegraphics[width=90mm]{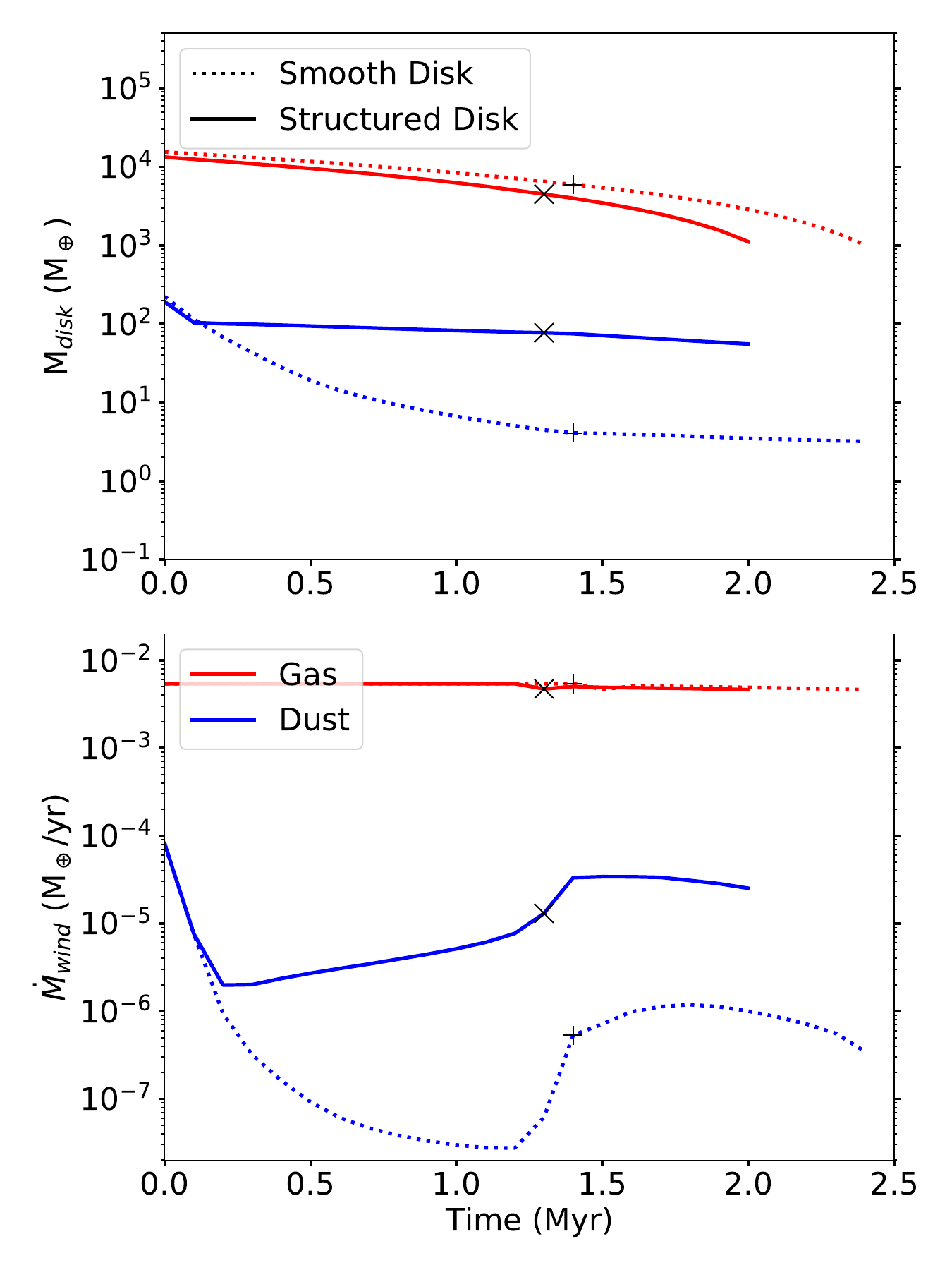}
 \caption{\textit{Top:} Mass evolution of the disk mass in gas (red lines) and dust (blue lines). \textit{Bottom:} Evolution of the mass loss rate of gas and dust by photoevaporative winds. 
 The markers indicate the moment when photoevaporation opens a cavity in the inner disk (\sgquote{+} for the initially smooth disk, \sgquote{x} for the initially structured disk).
 }
 \label{Fig_MassEvolution}
\end{figure}

\begin{figure}
\centering
\includegraphics[width=90mm]{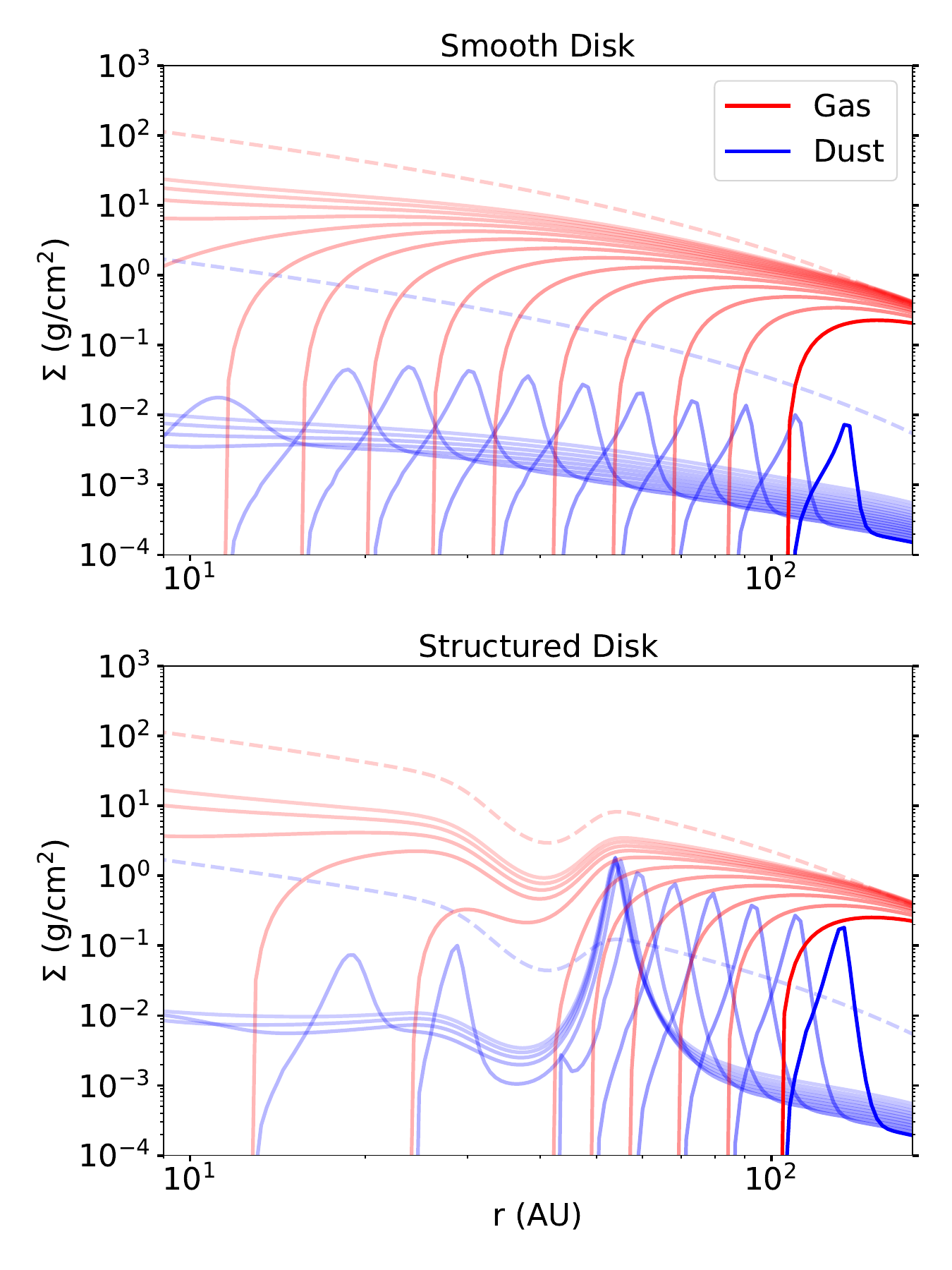}
 \caption{Evolution of the surface density profiles disks from $t = \SI{1}{Myr}$, and plotted every \SI{0.1}{Myr} (solid lines, with line opacity increasing with time). The dust surface density accounts for all the grain sizes.
 The initial condition is shown with dashed lines.}
 \label{Fig_SurfaceDensity}
\end{figure}

\begin{figure}
\centering
\includegraphics[width=95mm]{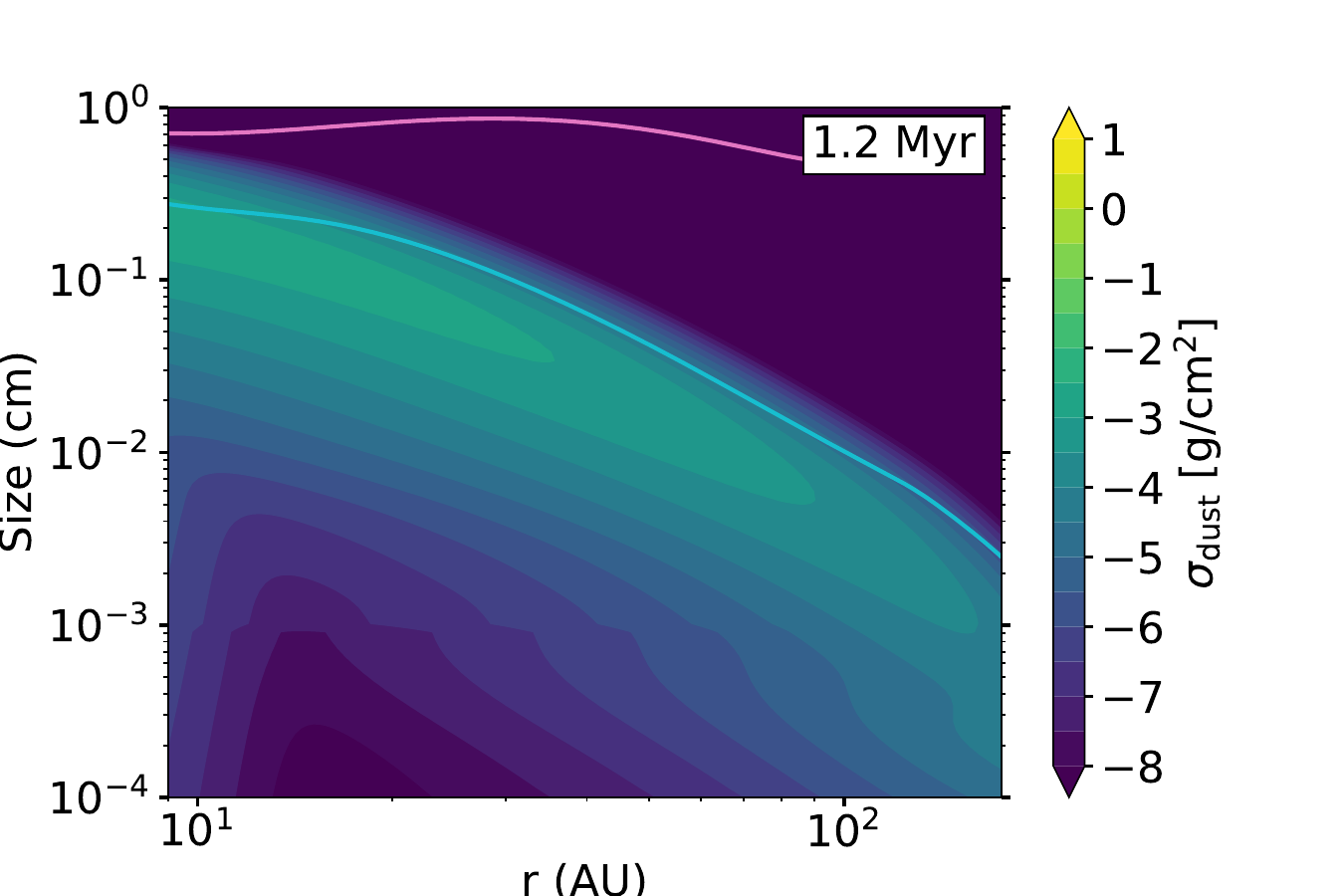}
\includegraphics[width=95mm]{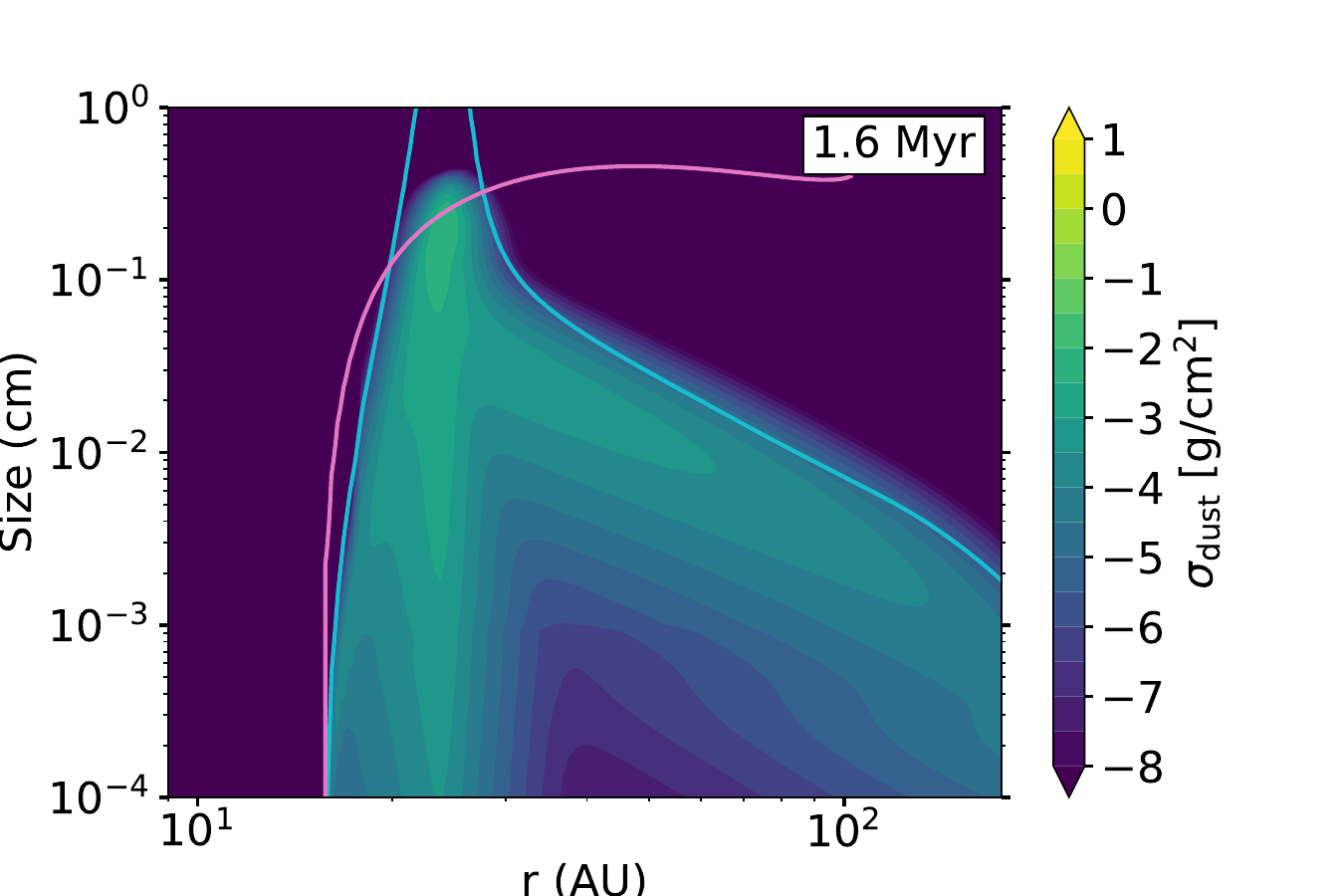}
\includegraphics[width=95mm]{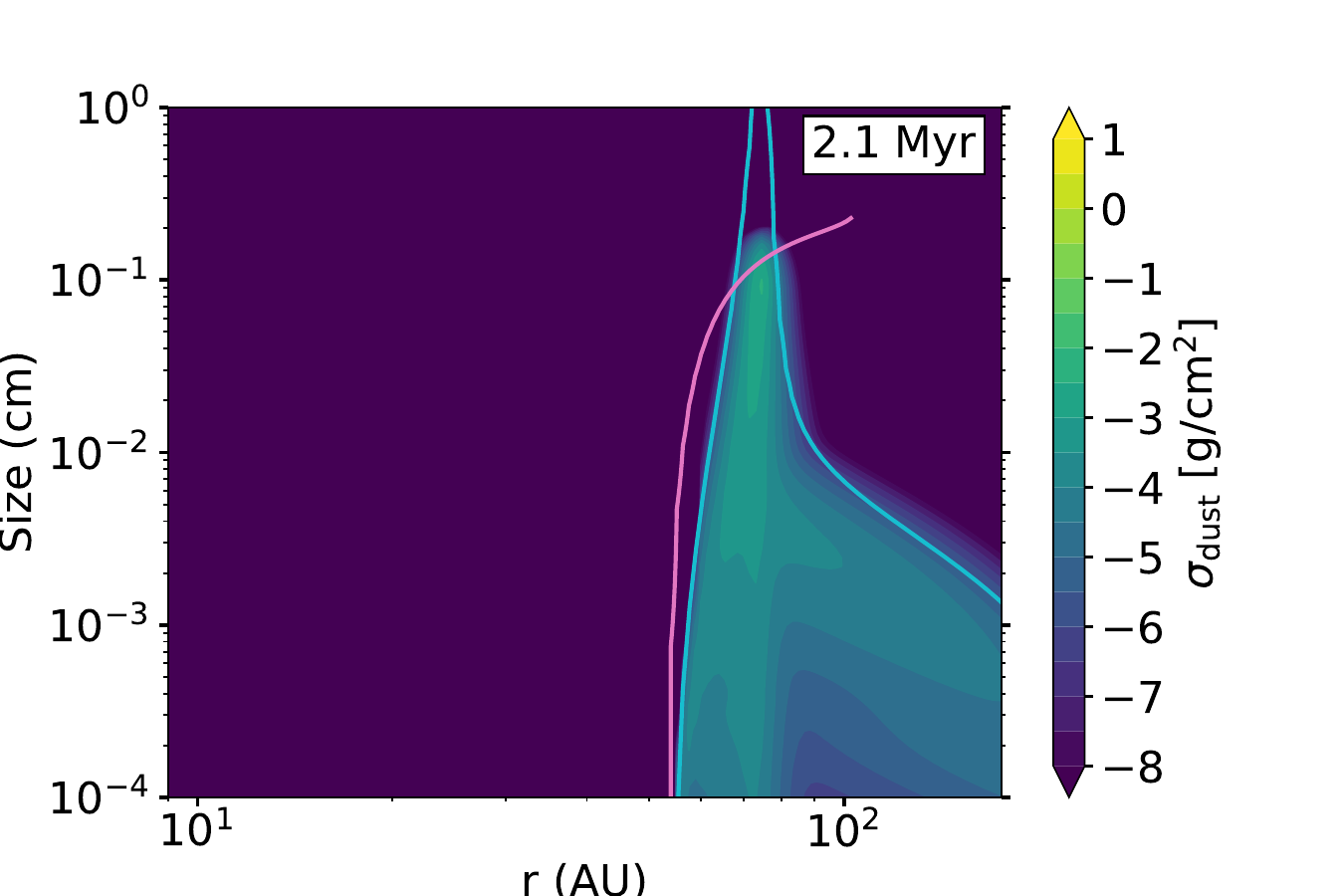}
 \caption{Dust size distribution for a smooth disk at \SI{1.2}{}, \SI{1.6}{}, and \SI{2.1}{Myr}. The drift (\textit{cyan}) and fragmentation (\textit{pink}) growth limits are also indicated.
 }
 \label{Fig_DustSizeDist_Smooth}
\end{figure}

\begin{figure}
\centering
\includegraphics[width=95mm]{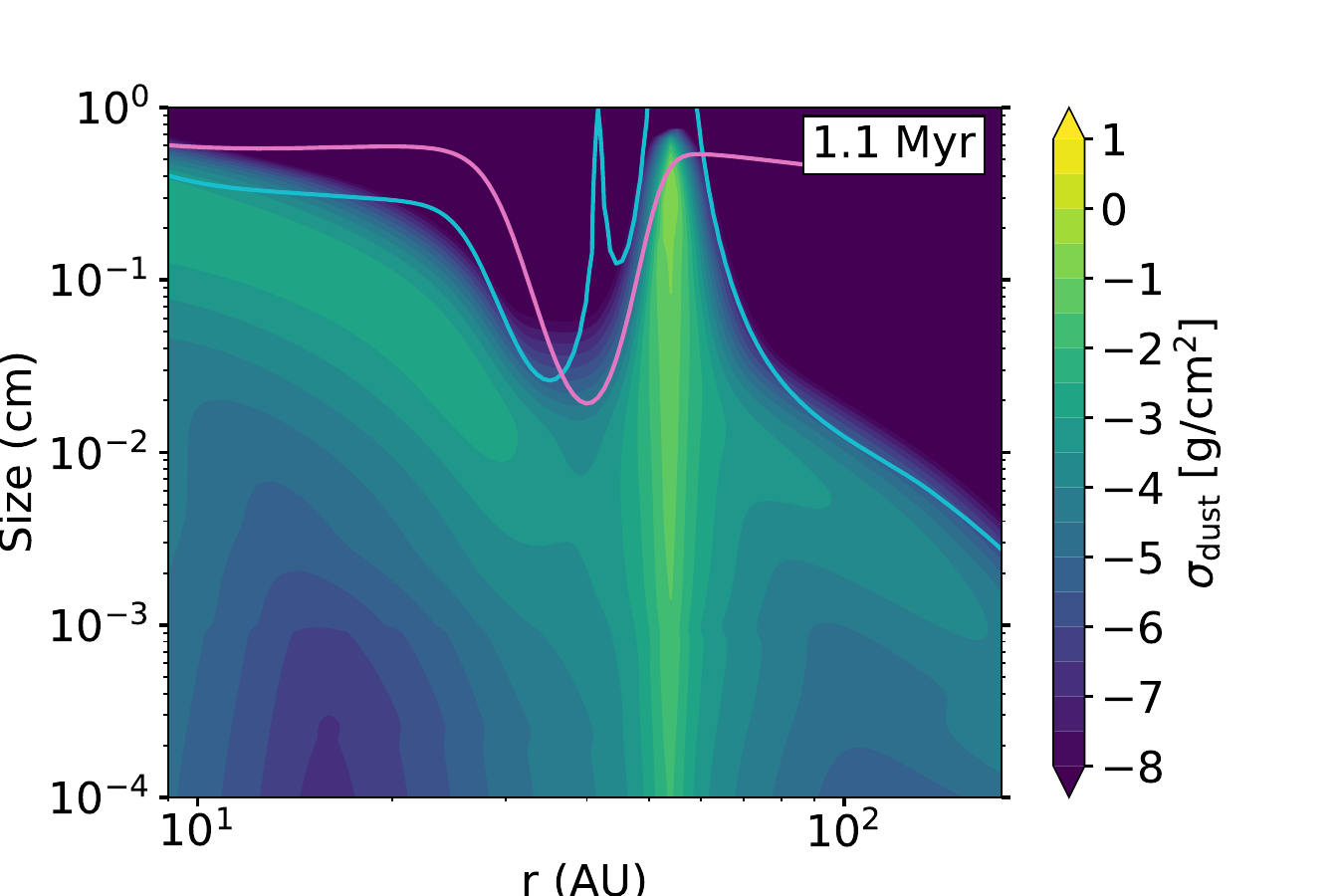}
\includegraphics[width=95mm]{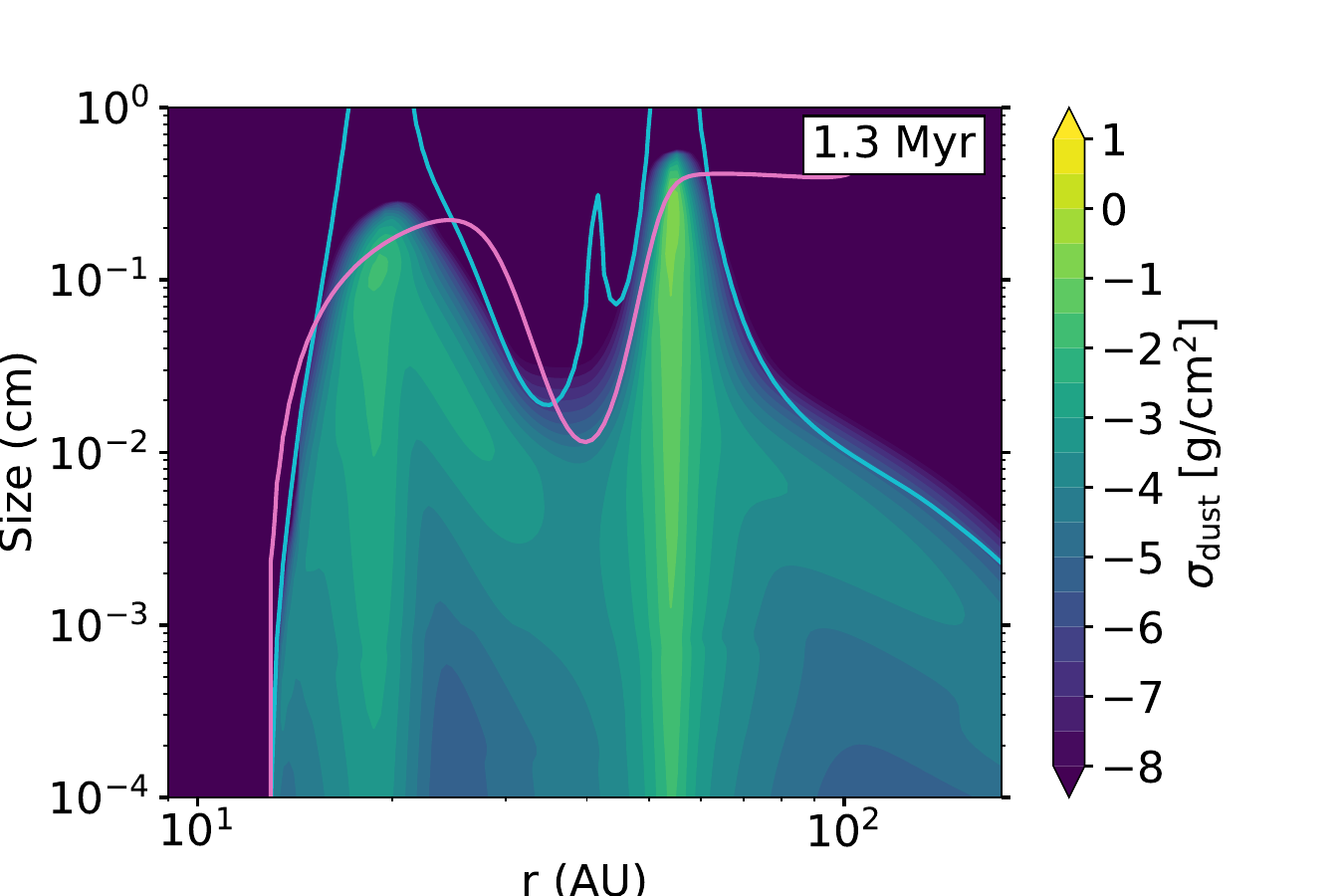}
\includegraphics[width=95mm]{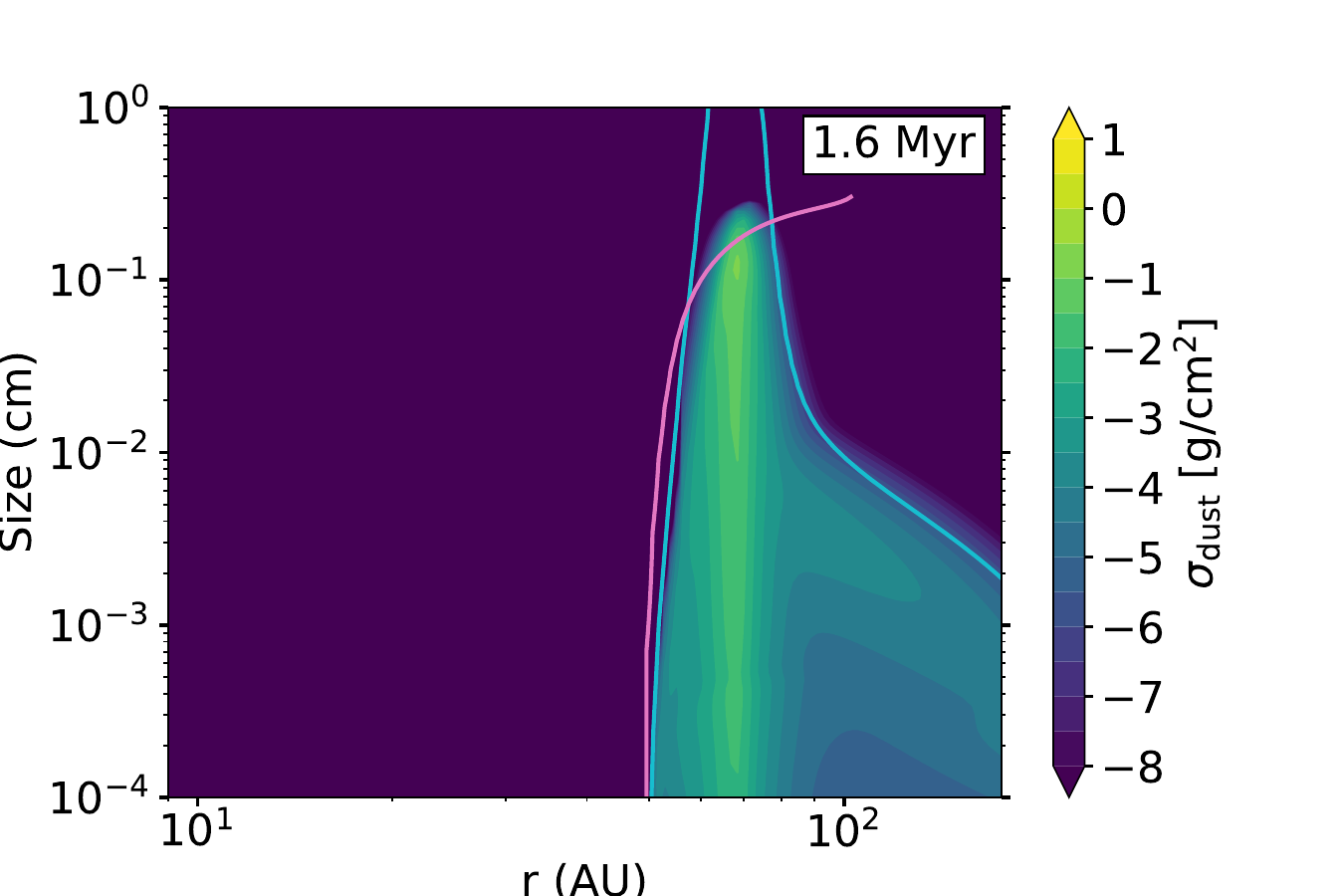}
 \caption{Dust size distribution for a structured disk at \SI{1.1}{}, \SI{1.3}{}, and \SI{1.6}{Myr}. The drift (\textit{cyan}) and fragmentation (\textit{pink}) growth limits are also indicated. The last two snapshots were selected to match the ones of \autoref{Fig_DustSizeDist_Smooth} in terms of the photoevaporative cavity size.
 }
 \label{Fig_DustSizeDist_Trap}
\end{figure}

In this section we present our results for the evolution of photoevaporating disks, focusing on the effect that early substructures (represented through a primordial gap in the gas component) have on the dust component, in terms of the distribution of solids and the corresponding observable quantities (millimeter fluxes and SEDs).\par

We often refer to the disk without the primordial gap as a \dbquote{smooth disk}, and the disk with the primordial gap as a \dbquote{structured disk}. We also distinguish between the \dbquote{gap} structure that is created through the variation in the $\alpha$ viscosity profile (\autoref{eq_alpha_bump}), and the \dbquote{cavity} that is carved by photoevaporative dispersal.

\subsubsection{Evolution of the dust distribution}
A disk with a gap-like substructure can efficiently trap dust grains at the local pressure maximum, so long as the substructure forms before any significant radial drift occurs \citep{Stadler2022}.
In contrast, in a disk without substructure the dust drifts very efficiently toward the star \citep[][]{Birnstiel2010, Pinilla2012_b, Pinilla2020}.
Our fiducial models show that, by the time that photoevaporation starts clearing the gas component from the inside out, after $\sim \SI{1}{Myr}$ of disk evolution (for $L_x = \SI{e30}{erg\, s^{-1}}$), the structured disk retains a higher mass of solids than the smooth disk, $\SI{77}{M_\oplus}$ and $\SI{4}{M_\oplus}$ respectively, out of the initial $\sim \SI{200}{M_\oplus}$ (see \autoref{Fig_MassEvolution}, top panel).\par

Once photoevaporation opens up a cavity in the inner regions, further dust drift toward the star is completely halted, since the edge of the cavity is also a pressure maximum that can trap solids (see the evolution of the surface densities, \autoref{Fig_SurfaceDensity}). 
From this point onward, the dust loss is driven exclusively by the entrainment in the photoevaporative winds, though additional loss terms such as planetesimal formation and removal by radiation pressure are neglected in our model \citep[][see discussion in \autoref{sec_Discussion_RelicDisks}]{Stammler2019, Owen2019}. \par

\autoref{Fig_MassEvolution} (bottom panel) shows that the dust loss rate by photoevaporative entrainment increases between one and two orders of magnitude after the photoevaporative cavity opens. 
One reason is that the gas loss rate locally increases at the edge of the photoevaporative cavity, where the material is directly irradiated by the central star. In \autoref{Fig_LossRateProfile} we see how the gas loss profile changes in this \dbquote{open cavity} scenario, with a sharp spike at the location of the cavity edge \citep[see also][their Eq. 4]{Picogna2019}.
The other reason, and perhaps more importantly, is that dust growth becomes limited by fragmentation around the pressure maxima which replenishes the population of small grains that are more easily entrained with the wind, and the gas surface density is also reduced at the photoevaporative cavity edge, sharply reducing the maximum grain size.
In contrast, the growth in the gas rich regions with steeper pressure gradients is limited only by drift, which results in a dust distribution dominated by large particles (this can bee seen from the grain size distributions in \autoref{Fig_DustSizeDist_Smooth} and \ref{Fig_DustSizeDist_Trap}), that are not easily entrained with the photoevaporative wind.\par

After the photoevaporative cavity opens, the remaining dust mass decreases from $\SI{77}{M_\oplus}$ to $\SI{55}{M_\oplus}$ for the structured disk model, and from $\SI{4}{M_\oplus}$ to $\SI{3}{M_\oplus}$ for the smooth disk model. 
These values imply that the total dust loss across the disk lifetime (or at least until the cavity size reaches $r_\textrm{cavity} = \SI{120}{AU}$ in our model) is mostly dominated by drift during the early stages of disk evolution, rather than entrainment in the photoevaporative winds \citep[see also][]{Ercolano2017_b}.\par


From the surface density profiles (\autoref{Fig_SurfaceDensity}) and the grain size distributions (\autoref{Fig_DustSizeDist_Smooth} and \ref{Fig_DustSizeDist_Trap}), we find that once a photoevaporative cavity opens, the remaining solid material is dragged along with the cavity outer edge, following the moving pressure maximum. 
For the smooth disk, this leads to the formation of a single dust trap at \SI{1.4}{Myr}, that moves outward as time passes. For the structured disk, on the other hand, we find that between \SI{1.2}{} and \SI{1.4}{Myr} there are two traps present, one that follows the photoevaporative cavity, and the other at the outer edge of the primordial gap, which should lead to a distinct disk morphology featuring two rings.
Eventually, both dust traps merge into one when the cavity catches up with the gap location, which then continues to move outward. We infer that the two ring morphology is more likely to be observed if the primordial dust trap is located at larger radii than the photoevaporative cavity opening radii. The latter would delay the merging of the two rings and increase the window of observation, though to get an accurate estimate of the likelihood of observing this evolutionary stage, a population synthesis model would be required. 
We also note that if a dead zone is present in the inner disk, the photoevaporative cavity opening radius can be located beyond \SI{10}{} to \SI{20}{AU} \citep{Garate2021}, meaning that disks with primordial dust traps located inside the dead zone radius (such as Jupiter's current orbit) would not lead to the described two ring morphology.\par
This particular behavior in the evolution of structured disks leads to a degeneracy between the properties of an observed dust ring and its potential origin, which is of particular interest for the study of transition disks, and we discuss more about it in \autoref{sec_Discussion_Degeneracy}.\\


\subsubsection{Millimeter emission and SED}

\label{sec_Results_FmmSED}
\begin{figure}
\centering
\includegraphics[width=90mm]{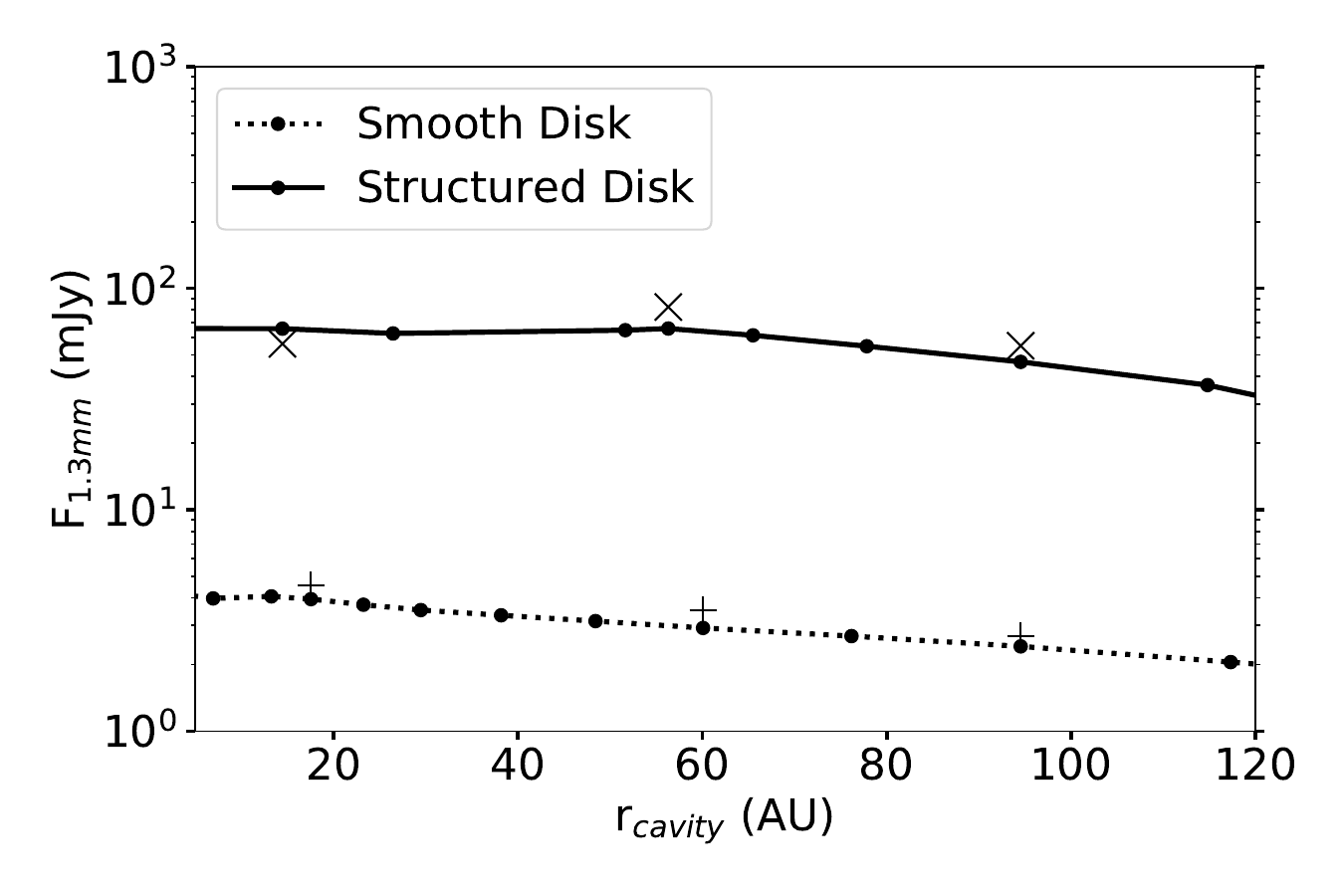}
 \caption{Millimeter fluxes $F_\textrm{mm}$, as a function of the size of the photoevaporative cavity for the smooth (dotted) and structured (solid) disk models. The solid and dotted lines represent flux from the optically thin approximation (\autoref{eq_Flux_OpticallyThin}). The markers are the fluxes obtained with RADMC-3D (\sgquote{+} for the smooth disk, \sgquote{x} for the structured disk). The disks are assumed to be at \SI{140}{pc}, and no inclination.
 The cavity size measurement is based on the dust distribution of millimeter sized grains.
 }
 \label{Fig_TotalIntensity}
\end{figure}

\begin{figure}
\centering
\includegraphics[width=90mm]{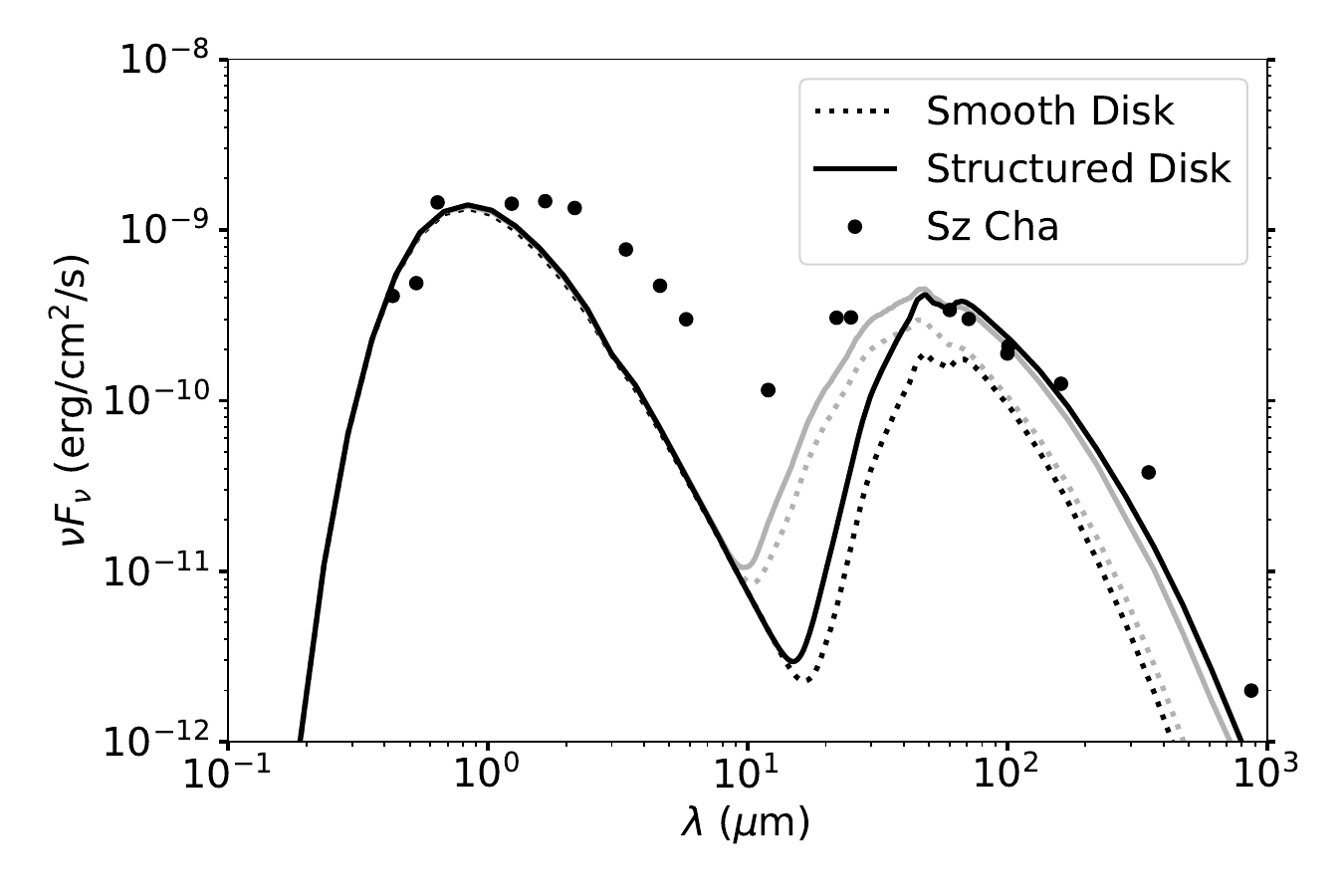}
 \caption{SEDs for the smooth (dotted) and structured (solid) disk models, when the photoevaporative cavity size  is $r_\textrm{cavity} \approx \SI{15}{AU}$ (gray), and $r_\textrm{cavity} \approx \SI{100}{AU}$ (black), assuming a distance of $\SI{140}{pc}$ and face-on.
 The data points show the SED of SzCha from \citet{vanderMarel2016} survey, re-scaled to a distance of $\SI{140}{pc}$.
 }
 \label{Fig_SED}
\end{figure}

Because the dust masses between the structured and smooth disks differ by over an order of magnitude during the photoevaporative dispersal, this also results in a similar difference between their corresponding luminosities in the millimeter continuum (\autoref{Fig_TotalIntensity}).
In our models, the flux of the structured disk is $F_\textrm{mm} \approx \SI{65}{mJy}$ (obtained from \autoref{eq_Flux_OpticallyThin}) by the time the cavity opens, and remains approximately constant until the cavity reaches the location of the primordial dust trap at $r \approx \SI{50}{AU}$, afterward the flux continues to decrease, reaching $\SI{44}{mJy}$ by the time the cavity has grown to $\SI{100}{AU}$. The smooth disk flux simply decreases approximately from $\SI{4}{mJy}$ to $\SI{2}{mJy}$.\par

We also note that the radiative transfer calculations with \texttt{RADMC-3D} differ only by a small factor from the values obtained with the vertical slab approximation when using the DSHARP opacity model. The difference in both fluxes is likely due the direct heating of the photoevaporative cavity edge by the stellar irradiation, and the proper treatment of the scattering and optical depth.\par


Finally, the SEDs (\autoref{Fig_SED}) show a similar behavior for both the smooth and structured disk, where the deficit in the NIR to MIR wavelengths becomes more prominent as the cavity size grows, and small grains are removed from the inner regions, meaning that our models would be classified as transition disks by their SED \citep[][]{Espaillat2014}. 
Additionally, the disks in our model display a high emission in the far infrared (FIR, around \SI{100}{\mu m}) which are comparable to those of the transition disk Sz-Cha \citep[][with the distance rescaled to 140 pc]{vanderMarel2016, Gaia3}, though we remark that this is not intended to be a representative comparison with the transition disk population.
In \autoref{sec_Discussion_RelicDisks} we discuss the implications of the FIR excess in the context of the relic disk problem.\par

\subsection{Effect of the X-ray luminosity}

\begin{figure}
\centering
\includegraphics[width=90mm]{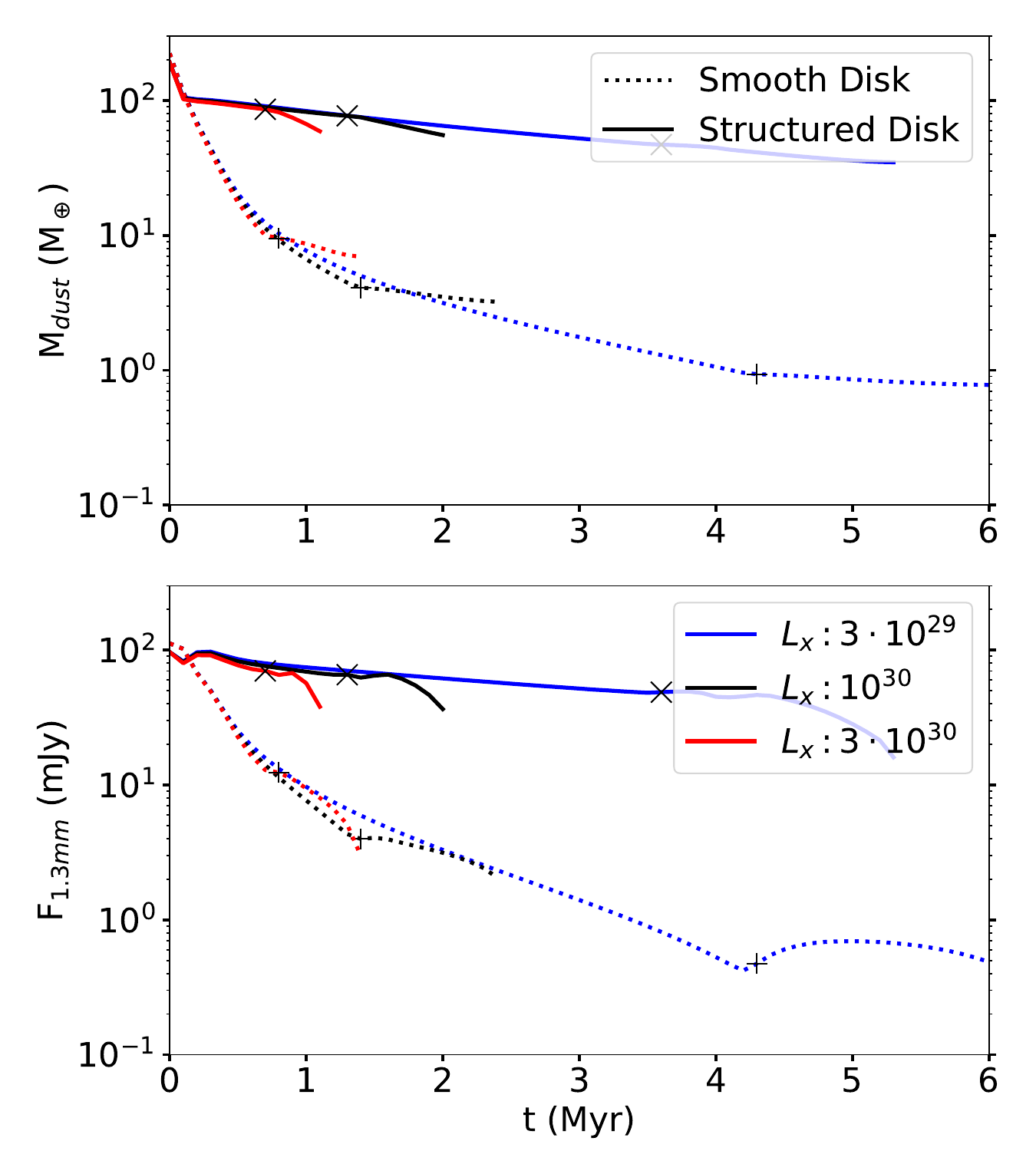}
 \caption{Evolution of the dust mass (\textit{top}) and disk flux at $\lambda = \SI{1.3}{mm}$ (\textit{bottom}, assuming a distance of \SI{140}{pc}), for different X-ray luminosities $L_x$, with the black line corresponding to the fiducial value.
 The markers indicate the moment when photoevaporation opens a cavity in the inner disk (\sgquote{+} for the smooth disk, \sgquote{x} for the structured disk).
 }
 \label{Fig_LxParameter}
\end{figure}

In this section we test the impact that different X-ray luminosities have on the evolution of the disk mass in the dust component, and on the corresponding flux in the millimeter continuum (see \autoref{Fig_LxParameter}). For the structured disk, we use again the fiducial gap amplitude of $A_\textrm{gap} = 4$ located at $r_\textrm{gap} = \SI{40}{AU}$.\par

We notice that for higher X-ray luminosities the photoevaporative cavity opens earlier (due to the higher mass loss rates), and that both the dust mass and millimeter flux are also higher when the photoevaporative cavity opens. This occurs because the dust drift, dust diffusion, and wind entrainment processes had less time to remove solid material from the disk.\par

The difference can be clearly seen in the mass and flux evolution of the smooth disk, where dust drift can only be stopped by the pressure maximum corresponding to the photoevaporative cavity.
For the highest X-ray luminosity ($L_x = \SI{3e30}{erg\, s^{-1}}$) the millimeter flux of the smooth disk is $F_\textrm{mm} = \SI{12}{mJy}$ (at \SI{0.8}{Myr}, when the cavity opens), while for the lowest X-ray luminosity ($L_x = \SI{3e29}{erg\, s^{-1}}$) the flux is only  $F_\textrm{mm} = \SI{0.4}{mJy}$ (at \SI{4.3}{Myr}).\par

On the other hand, while the structured disk can trap dust particles at the local pressure maximum, some of them will still diffuse through the gap and be lost to the star. 
We find that dust entrainment only accounts for a minor fraction of the dust removal in structured disks before the opening of the photoevaporative cavity, with rates between $\SI{e-6}{M_\oplus\, yr^{-1}}$ and $\SI{e-5}{M_\oplus\, yr^{-1}}$ (for $L_x = \SI{3e29}{erg\, s^{-1}}$ and $\SI{3e30}{erg\, s^{-1}}$, respectively).
The millimeter flux of the structured disk is $F_\textrm{mm} = \SI{70}{mJy}$ when the cavity opens (at \SI{0.7}{Myr}) for the highest X-ray luminosity, and $F_\textrm{mm} = \SI{49}{mJy}$ for the lowest X-ray luminosity (at \SI{3.6}{Myr}).
We expect that disks with multiple dust traps would be able to retain more material and prevent further dust loss due to diffusion \citep[][]{Pinilla2012_b, Pinilla2020}.\par

Another feature that we observe for each pair of simulations with the same X-ray luminosity, is that the inner cavity opens earlier in the structured disk than in the smooth one by \SI{0.2}{Myr}, a behavior that is also seen in the simulations of \citet{Rosotti2013} when the effects of photoevaporation and planet-disk interactions are considered. 
This occurs because the presence of the gap structure seems to speed up the viscous evolution of the disk by a small factor, reducing the gas accretion rate in the inner regions faster, and allowing for photoevaporative dispersal to start earlier. This is also the reason why the gas mass decreases slightly faster in the initially structured disk than in the initially smooth disk shown in \autoref{Fig_MassEvolution} (top panel).

\subsection{Effect of the trap location and amplitude}
\begin{figure}
\centering
\includegraphics[width=90mm]{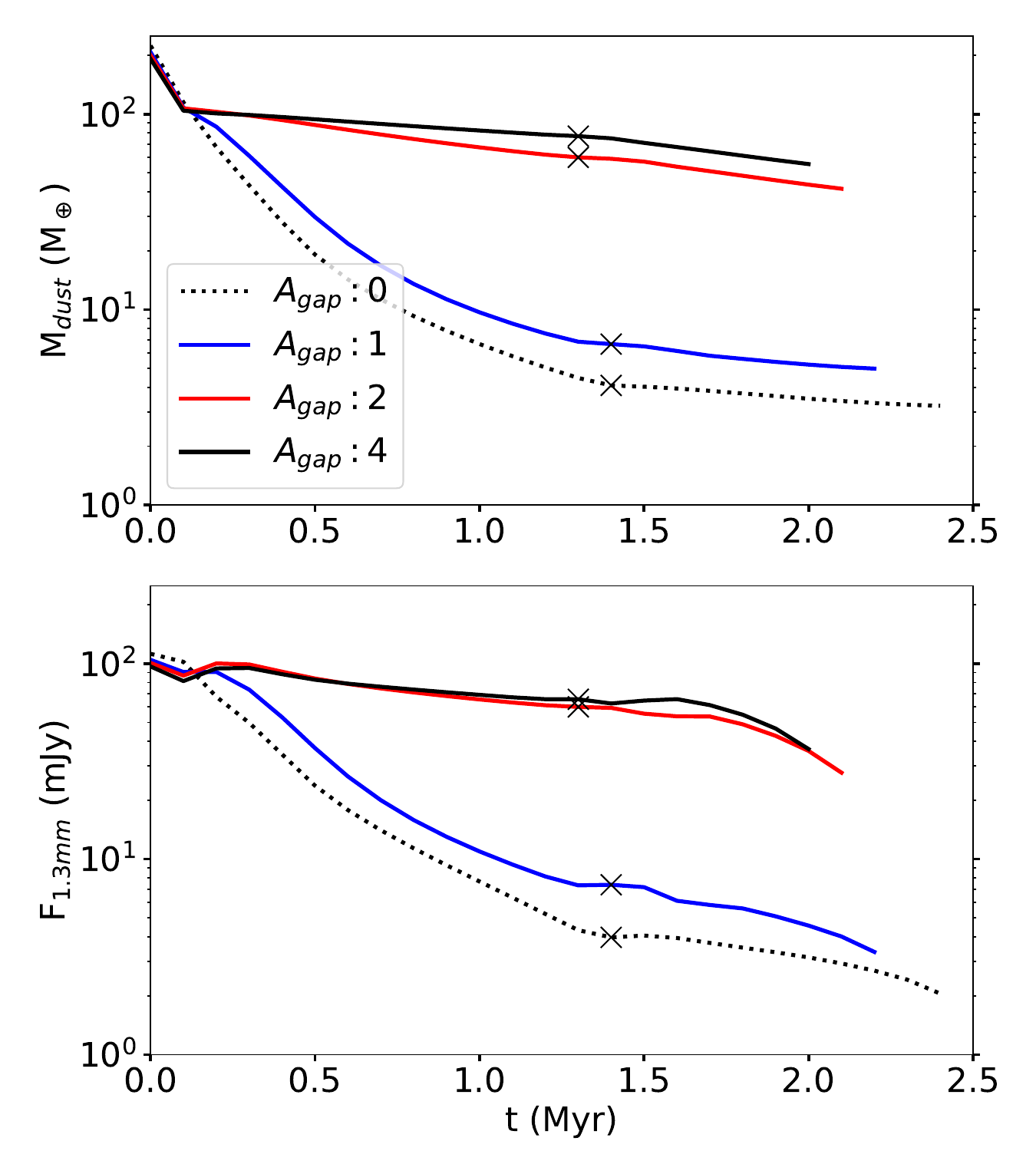}
 \caption{Same as \autoref{Fig_LxParameter}, but for structured disks with different gap amplitudes, with a fixed location at $r_\textrm{gap}=\SI{40}{AU}$ and X-ray luminosity of $L_x = \SI{e30}{erg\, s^{-1}}$. Notice that the axis scales are different from \autoref{Fig_LxParameter}.
 }
 \label{Fig_AgapParameter}
\end{figure}

\begin{figure}
\centering
\includegraphics[width=90mm]{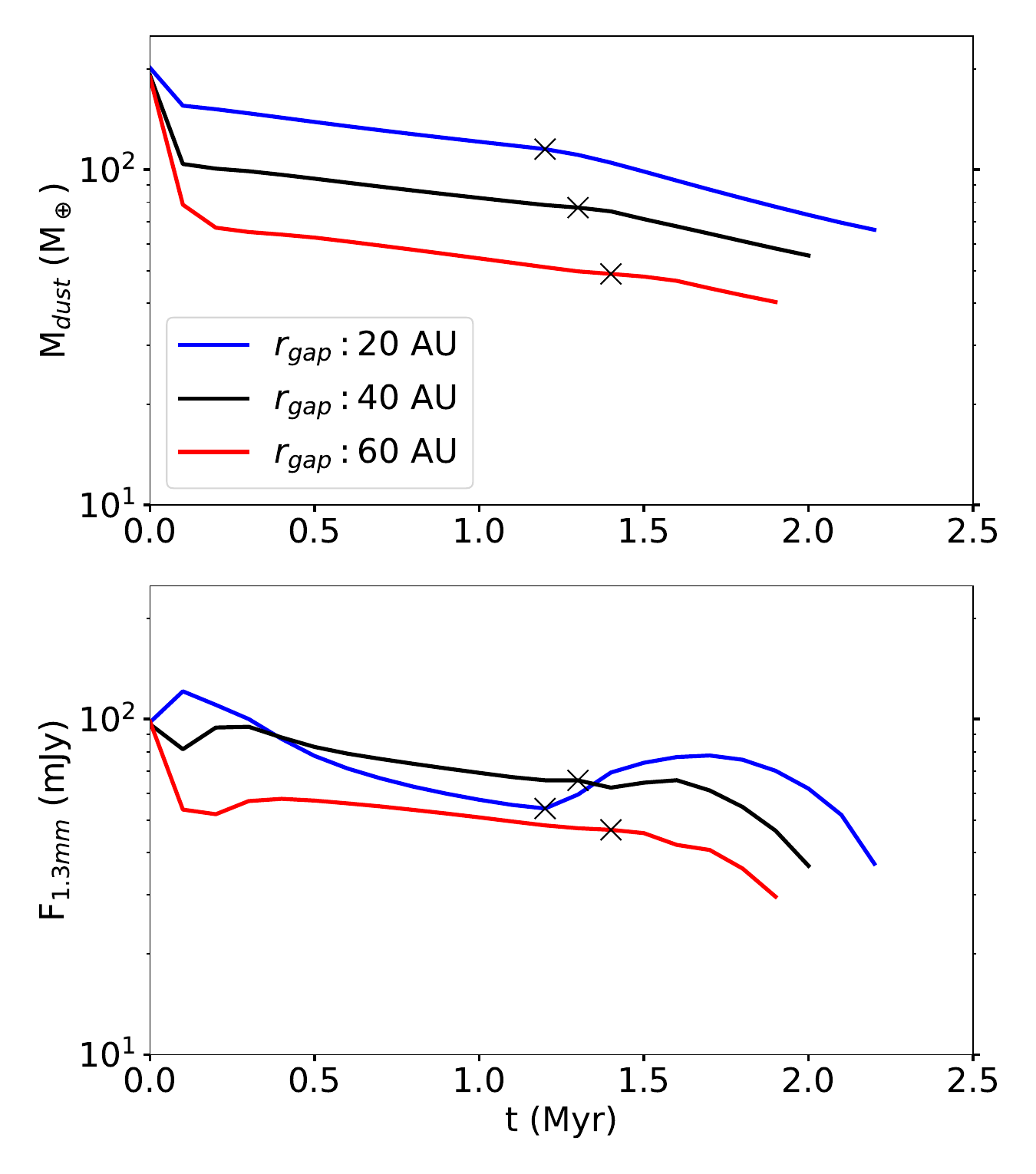}
 \caption{Same as \autoref{Fig_LxParameter}, but for structured disks with different gap locations, with a fixed amplitude of $A_\textrm{gap}=4$ and X-ray luminosity of $L_x = \SI{e30}{erg\, s^{-1}}$. 
 Notice that the y-axis scale is different from \autoref{Fig_AgapParameter}.
 }
 \label{Fig_RgapParameter}
\end{figure}

In this section we test the impact that the gap amplitude and location have on the evolution of the mass in the dust component and the respective flux in the millimeter continuum (\autoref{Fig_AgapParameter}  and \ref{Fig_RgapParameter}). For the X-ray luminosity we use the fiducial value of $L_x = \SI{e30}{erg\, s^{-1}}$.\par

We observe that a minimum amplitude of $A_\textrm{gap} = 2$, which roughly corresponding to a Saturn mass in our models, seems to be required for a gap at $r_\textrm{gap} = \SI{40}{AU}$ to effectively trap the dust particles that drift from the outer region (\autoref{Fig_AgapParameter}, top panel). The dust mass for this disk, when the inner cavity opens, is about $\SI{60}{M_\oplus}$, in comparison with the $\SI{77}{M_\oplus}$ measured for the disk with the deeper gap ($A_\textrm{gap} = 4$).
In contrast, the disk with $A_\textrm{gap} = 1$ is unable to stop the dust drift and is almost indistinguishable from the completely smooth disk, retaining a dust mass of only $\SI{7}{M_\oplus}$ at the cavity opening.\par

The location of the gap (relative to the disk characteristic radius $r_c$) on the other hand determines the size of the dust reservoir that can be retained at a given dust trap. Because dust grains tend to drift inward, all the dust that is initially inside the gap location is rapidly lost into the star, as can be seen during the first $\SI{0.1}{Myr}$ of disk evolution (\autoref{Fig_RgapParameter}, top panel). Only the dust grains that are further away than the local pressure maximum can be potentially trapped in it (if we ignore diffusion and wind entrainment), and therefore gaps that are closer to the star result in a higher dust content during the disk evolution and dispersal. For comparison, the dust mass of the disk with the innermost gap at $\SI{20}{AU}$, is approximately double the mass of the disk with the outermost gap at $\SI{60}{AU}$, at the moment of the cavity opening.\par

In terms of the millimeter flux we notice that the disk with $r_\textrm{gap} = \SI{20}{AU}$ is slightly fainter than the fiducial model with $r_\textrm{gap} = \SI{40}{AU}$ upon the opening of the inner cavity. 
This might sound counter-intuitive, since the simulation with $r_\textrm{gap} = \SI{20}{AU}$ has a larger dust mass (\autoref{Fig_RgapParameter}) at all times. However, we also notice that a dust trap located further inside into the disk will have both a higher optical depth ($\tau_\nu \gtrsim 1$) and a smaller surface area, where both of these effects would then contribute to reduce the total flux observed (see \autoref{eq_Flux_OpticallyThin}).
In other words, at the moment of the cavity opening the disk with the innermost dust trap has more solid material, but this material is \dbquote{hidden} from the observer.
As the cavity starts growing and the dust trap moves out, the disk with the initially innermost gap becomes the brightest, reaching $F_\textrm{mm} = \SI{78}{mJy}$ at its peak (after the inner cavity opens).\par

We note that a disk with multiple traps located in the inner and outer regions should be able to retain a higher fraction of the initial dust mass, and display higher fluxes in the millimeter continuum.
We also want to point out that the ratio between the gap location and the disk initial size ($r_\textrm{gap}/r_c$) should be more important to predict the amount of dust trapped than their absolute value, since that the same values of $r_\textrm{gap}$ presented in this section may trap more (or less) solid material for disks that are initially more extended (or compact).

\subsection{Comparison between opacities}\label{Sec_Results_Opac}
\begin{table}
 \caption{Flux comparison between \citet{Birnstiel2018} and \citet{Ricci2010} opacity models, at $\lambda = \SI{1.3}{mm}$.}
 \label{Table_OpacityCompare}
 \centering
  \begin{tabular}{l c | c  c}
    \hline \hline
    \noalign{\smallskip}
    Disk Model & $t$ [Myr] & $F_\textrm{Birnstiel}$ [mJy] & $F_\textrm{Ricci}$ [mJy]\\
    \hline
    \noalign{\smallskip}
           & 1.6 & 4.6 & 18.2 \\
    Smooth & 2.1 & 3.5 & 14.7 \\
           & 2.3 & 2.7 & 11.9 \\
    \hline
    \noalign{\smallskip}
               & 1.3 & 56.1 & 114.0 \\
    Structured & 1.6 & 82.3 & 166.1 \\
               & 1.9 & 55.0 & 169.6 \\
    \hline
  \end{tabular}
\end{table}

Due to the uncertainty in the opacities of dust grains in protoplanetary disks, particularly in the amount of carbonaceous material, we chose to perform an additional set of radiative transfer calculations, this time with the opacity model of \citet{Ricci2010}, and compare the corresponding millimeter fluxes against those obtained with the \citet{Birnstiel2010} (DSHARP) model in \autoref{sec_Results_FmmSED}.\par

From \autoref{Table_OpacityCompare} we observe that the flux in the millimeter continuum  is always higher when the \citet{Ricci2010} opacity model is considered, which is to be expected since the absorption opacities are also higher.
The trend between smooth and structured disks is maintained no matter the opacity model used, with the smooth disks being fainter ($F_\textrm{mm} < \SI{20}{mJy}$) and the structured disks being brighter ($F_\textrm{mm} > \SI{110}{mJy}$).\par

Finally, we notice that the relative increase in flux when using the \citet{Ricci2010} opacities is higher in the smooth disks (a factor between $4$ to $5$) than in the structured disks (a factor between $2$ to $4$). This occurs because the structured disks have regions that are optically thick (at the dust traps, for example), while on the other hand most of the smooth disks are optically thin, and therefore more sensitive to changes in the opacity model.

\section{Discussion}\label{sec_Discussion}
\subsection{Explaining the observed transition disks with photoevaporation models} \label{sec_Discussion_TransitionDisks}
Since first discovered, transition disk properties have remained a challenge to theoretical models.
These objects display deep cavities in the dust component, as probed by the deficit in NIR and MIR emission \citep{Strom1989, Skrutskie1990, vanderMarel2016} and resolved continuum observations \citep[][]{Andrews2011, pinilla2018, Francis2020},  while displaying relatively high accretion rates ($\dot{M}_\textrm{acc} \sim \SI{e-10}{}$ to $\SI{e-8}{M_\odot\, yr^{-1}}$) that indicate the presence of long lived inner disks \citep{Cieza2012, Alcala2014, Manara2017}. Additionally, transition disks seem to be distributed across a mm-faint ($F_\textrm{mm} < \SI{30}{mJy}$) and a mm-bright ($F_\textrm{mm} > \SI{30}{mJy}$) population, according to their fluxes in the millimeter continuum \citep[see the review by][]{Owen2016_review}. However, it should be noted that the faint transition disks, which are identified based on their SEDs, could be instead highly inclined disks rather than actual transition disks with cavities \citep{vanderMarel2022}. \par

Giant planets have often been proposed as an explanation for transition disk properties, since these can carve deep gaps in the millimeter continuum by trapping millimeter grains, while still allowing for the flow of gas toward the inner disk \citep[][]{dong2012, Pinilla2012, ovelar2013, Owen2014}. On that line, a recent study of \citet{vanderMarel2021} proposes that the population of observed transition disks can be linked to the population of detected exoplanets, under the assumption that planets more massive than Jupiter are responsible for the transition disk substructures, and that these planets then migrate inward.\par

Yet, despite the high masses predicted for the planetary companions \citep[Jupiter-sized or larger,][]{Muley2019, vanderMarel2021_b}, to this day the only disk with a confirmed planet detection is PDS-70 \citep[][]{Keppler2018}. While some of these hypothetical planets would still be beyond the current detection limits \citep[][]{Asensio2021}, we cannot discard that other mechanisms may be actually responsible for transition disks.\par

The simulations presented in this work \citep[and in combination with][]{Garate2021}, offer an additional pathway to explain the properties of transition disks without the presence of very massive planets, which would also explain why we have not detected these giants in the first place.
In \autoref{sec_Results} we showed that when substructures are present, the dust grains are retained in the local pressure maxima through the lifetime of the protoplanetary disk. 
Then, once the inside-out dispersal due to photoevaporation begins, all the trapped material is dragged along with the edge of the photoevaporative cavity, which results in a bright disk with a wide expanding cavity (\autoref{Fig_TotalIntensity}), consistent with the mm-bright population of transition disks.\par

Now, notice that the only requirements imposed on these substructures to explain transition disk properties of the mm-bright population would be a minimum amplitude (i.e., gap depth) to trap solid material, and dust traps located closer to the star favor would favor brighter disks (see \autoref{Fig_AgapParameter} and \ref{Fig_RgapParameter}).
This means that even planets with masses between those of Saturn and Jupiter, without imposing strong constraints on their location, could create the necessary substructures to reproduce a population of millimeter bright transition disk, under the assumption that internal photoevaporation is the primary dispersal process responsible for opening the cavity. \par

On the other hand, if substructures are absent, then dust drift depletes most of the solid material in the protoplanetary disk, which leads to a disk with a faint emission in the continuum and a cavity driven exclusively by photoevaporation, similar to the transition disks found in the mm-faint population.
We note that these analogies between smooth disks and mm-faint population, and between structured disks and mm-bright population, hold independently of the dust opacity (see \autoref{Sec_Results_Opac}).\par

In summary, our model suggests a synergy between photoevaporation and substructures, where photoevaporation is responsible for creating deep cavities in the gas and dust (both micron and millimeter sized components), while the presence or absence of early substructures determines the dust trapping and the disk millimeter flux, but not necessarily the observed cavity size. Without photoevaporation a moderate sized planet would not be able to create a deep cavity, and without early substructures photoevaporation would not be able to trap the amount of dust observed in bright transition disks.\par

Regarding the absolute dust mass and millimeter flux values found in our simulations for the smooth and structured disk models, we highlight that these depend on the initial dust mass used in the simulation setup, and it is more meaningful to look at the fraction of dust that was retained or lost relative to the initial value. For our fiducial setup that means that $\approx25\%$ of the initial dust mass was still present for the structured disk, and only $\approx2\%$ for the smooth disk. 
Transition disks with higher millimeter fluxes than those found in our current paper, such as LkCa 15 or GM Aur \citep[][respectively]{Facchini2020, Huang2020}, could be explained if we consider, for example, a disk that is initially more massive.\par

Another characteristic of transition disks are their relatively high accretion rates. Previously these seemed to be incompatible with photoevaporation models, which typically overpredicted the fraction of non-accreting disks with large cavities \citep[][]{Owen2011, Picogna2019}. However, the works of \citet{Morishima2012} and \citet{Garate2021} showed already that a dead zone in the inner regions leads to long-lived inner disks, capable of sustaining high accretion rates while photoevaporation opens a cavity in the outer disk, and the work of \citet{Ercolano2018} and \citet{Wolfer2019} suggests the fraction of accreting disks with photoevaporative cavities could be higher if the disks are relatively depleted in carbon and oxygen.\par

Based on the measurements of X-ray luminosities \citep[][]{Preibisch2005}, on the models of dead zones that suggest that these should be both common and stable \citep[][2022b, in prep.]{Delage2022}, and on the abundance of observed substructures in disks \citep[e.g.,][]{williams2011, Andrews2018}, which additionally are expected to form early in the disk evolution \citep{Stadler2022}, we can expect for all of the above-mentioned ingredients to influence evolution of protoplanetary disks. 
Then, in a combined disk model we would have that photoevaporation takes care of opening wide cavities, dead zone properties set the lifetime of the inner disk and accretion rate onto the star, and the dust trapping in early substructures (by a Saturn-to-Jupiter mass planet for example) determines the disk flux.\par

For example, our work shows that bright transition disks such as DM Tau, could still be explained through photoevaporation \citep[see Section 6.3 by][though a higher initial disk mass would be most likely required to explain this particular object]{Francis2022}, provided that there are additional dust trapping mechanisms acting during the early stages of disk evolution to explain the millimeter fluxes.
With this new information taken into account, we can relax the constraints on massive planets as the sole cause of transition disks with wide gaps, and also relax the requirements on planet migration to link the transition disk and exoplanet populations \citep[][]{vanderMarel2021}.

\subsection{The degeneracy of planet properties when photoevaporation is considered} \label{sec_Discussion_Degeneracy}
\begin{figure*}
\centering
\includegraphics[width = 180mm]{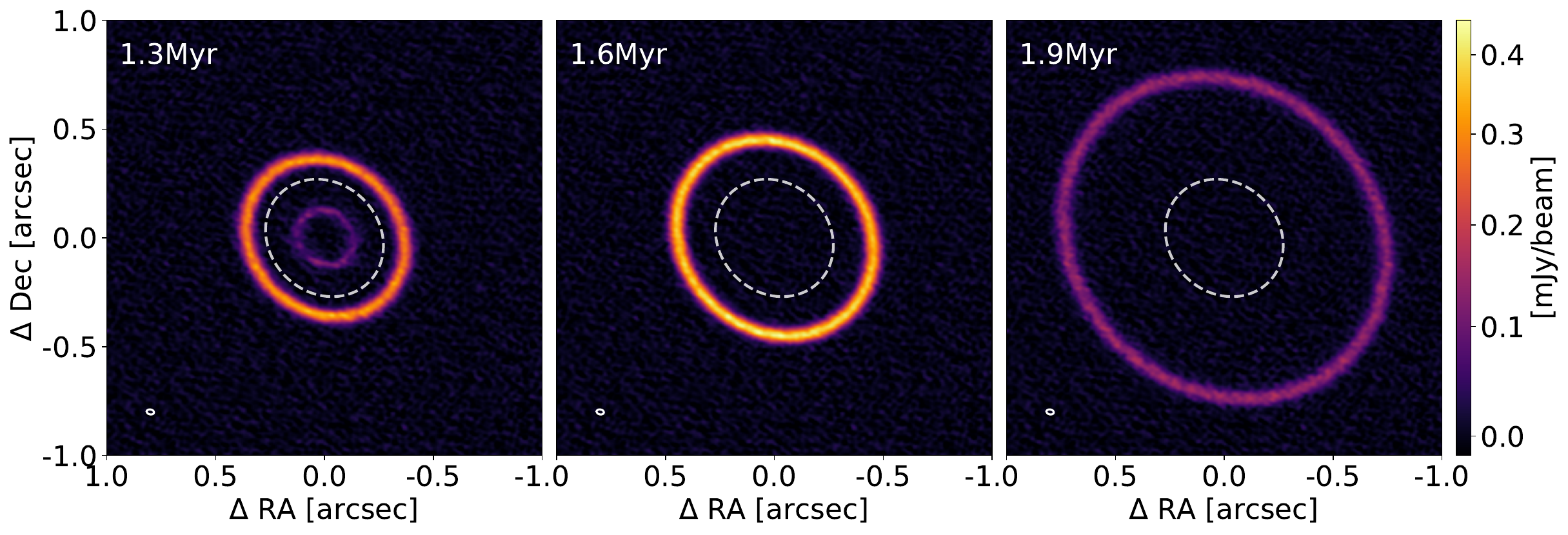}
 \caption{Synthetic ALMA observations at \SI{1.3}{mm} of the structured disk model at $\SI{1.3}{Myr}$, $\SI{1.6}{Myr}$, and $\SI{1.8}{Myr}$, generated using the \texttt{SIMIO} package (Kurtovic et al., in prep.) to post-process our radiative transfer model. 
 The image shows how our disk would look if it was observed with the same ALMA configuration of Elias24 \citep[][from the DSHARP sample]{Huang2018b}, assuming a distance of \SI{139}{pc} and an inclination of $29^\circ$. The beam size is plotted in the lower-left corner. 
 The orbit of the primordial gap  ($r_\textrm{gap} = \SI{40}{AU}$) is marked with a dashed line. We note that this is not intended to be a comparison \dbquote{with} Elias24.
 }
 \label{Fig_SIMIO}
\end{figure*}

In this section we want to emphasize an (unfortunate) consequence suggested by our model in terms of the characterization of planet candidates within transition disk cavities.
Our model shows a disk in which photoevaporative dispersal controls the size of the cavity that expands from the inside out, and primordial substructures (which in this subsection we assume to be caused by planets) control the dust trapping and millimeter flux.%
Here, we can distinguish two stages in the photoevaporative dispersal process. An early stage when the photoevaporative cavity is smaller than the planet orbit (which leads to the formation of two dust rings), and a later stage where the photoevaporative cavity size is larger than the planet orbit.
In the latter scenario we find that the size of the cavity is completely independent of the planet orbital location, and the disk flux in the millimeter continuum is only mildly sensitive to the gap amplitude and location (\autoref{Fig_AgapParameter} and \ref{Fig_RgapParameter}).\par

Therefore, our model suggest that there is to a strong degeneracy in which we cannot draw reliable constraints on the mass or location of a planet candidate within a transition disk based on the properties of the cavity.
A wide cavity with a bright ring in the millimeter continuum might be very well caused by a super-Jupiter mass planet near the cavity edge, or by photoevaporative dispersal and a Saturn mass planet hidden in the inner regions. In particular, a planet inside a photoevaporative cavity would be effectively disconnected from the rest of the protoplanetary disk, and would not display a detectable circumplanetary disk during the dispersal process, and this could in principle explain why we have not detected more planet companions or circumplanetary disks other than PDS70 \citep[][]{Keppler2018, benisty2021}.\par

To further illustrate this situation, we show a synthetic observation of our structured disk model in \autoref{Fig_SIMIO}, which was generated using the \texttt{SIMIO}\footnote{\href{https://www.nicolaskurtovic.com/simio}{www.nicolaskurtovic.com/simio}} package
(Kurtovic et al., in prep.) to post-process our radiative transfer models as if they were observed by ALMA with the template configuration of Elias24 \citep[][]{Huang2018b}.
The image sequence shows that once the photoevaporative cavity opens (\SI{1.3}{Myr}, left panel) two rings can be initially identified, where the inner one is caused by the dust trapped at the photoevaporative cavity edge, while the outer one is caused by the dust trapped outside the planet gap. 
In our simulations, the inner disk is fainter than the outer one (at this early stage) because the photoevaporative cavity was only able to trap the remaining material from the inner regions, while the primordial gap trapped most of the material available from the outer regions. However, this stage is short-lived and lasts only for $\sim \SI{0.2}{Myr}$  in our the fiducial setup.
As time passes, the photoevaporative cavity expands, which causes the inner dust ring to merge with the outer one (\SI{1.6}{Myr}, middle panel), and then to continue to expand well beyond the planet orbital location (\SI{1.9}{Myr}, right panel).
\autoref{Fig_SIMIO} suggests that by the last snapshot it would be impossible to infer anything about the planet location (that was responsible for the dust trapping) from the ring morphology and location in the millimeter continuum alone.\par

In order to distinguish if a cavity could be carved by photoevaporation or by a massive planetary companion, other types of signatures should be considered. 
Recent theoretical studies have focused on modeling the observational signatures from photoevaporative winds, including the expected dust content entrained with the gas \citep[][]{Franz2022a, Franz2022b, Rodenkirch2022}, which could be used to point toward the photoevaporative origin of some cavities in transition disks where a planet companion has not yet been found, though these models are strongly dependent on the dust reservoir at the cavity edge.
As shown in \autoref{Fig_SurfaceDensity}, the peak of $\Sigmadust$ is always at larger radii than that of $\Sigmagas$, which means the dust signature in the wind may be even fainter than predicted by \citet{Franz2022a}; and further studies investigate the correlation between the gas and dust distributions at the cavity edge in more detail \citep{Picogna2023}.\par

Local perturbations in the gas kinematics, for example, can be linked to a planetary companion embedded in the gap \citep[e.g.,][]{Perez2015, Pinte2019, Izquierdo2021}.
Asymmetries such as spirals would not be caused by photoevaporation, but other processes in addition to planetary companions can also cause them, such as self-gravitating instabilities \citep[e.g.,][]{Lodato2005, Meru2017} and shadows cast by an inclined inner disk \citep{Montesinos2018, Cuello2019}.\par

Characterizing the gas content inside the millimeter continuum cavities can also help to differentiate between the different scenarios \citep[][]{marel2016b}, since planets tend to carve deeper cavities in the dust than in the gas, while photoevaporation carves deep cavities in both of the components \citep[see reviews by][]{Owen2016_review, Ercolano2017_Review}. 
However, we note that the presence of a dead zone may also result in a long-lived inner disk inside the photoevaporative cavity \citep{Morishima2012, Garate2021}. This inner disk would be rich in gas and poor in dust, and would act as an accretion reservoir for the star while photoevaporation opens a cavity in the outer regions, effectively mimicking some of the features produced by a planet-carved gap.\par

With this section we have highlighted how different mechanisms can interact together to produce similar features to those of a very massive planet, and that this degeneracy should be taken into account in the cases where no further evidence of the expected planet candidates is found.


\subsection{On the relic disk problem} \label{sec_Discussion_RelicDisks}

Several explanations have been proposed to solve the overprediction of relic disks by photoevaporative models. Some studies focus on reducing the predicted fraction through faster dispersal processes, such as thermal sweeping \citep{Owen2013}, and low carbon and oxygen abundances \citep[][]{Ercolano2018, Wolfer2019}, or through long lived inner disks with dead zones that can sustain the high accretion rates for longer times \citep[][]{Garate2021}.
Another possibility is that the radiation pressure is very efficient at removing dust during photoevaporation dispersal, since the stellar photons can transfer their momentum directly to the solid particles, which can lead to very faint disks  that would be hard to detect in the IR \citep[][]{Owen2019}.\par

In this work we find that the SEDs from \autoref{Fig_SED} still display a high amount of FIR emission, which is above the median of class II disks from nearby star-forming regions \citep[][]{Ribas2017}, and in line with the high FIR luminosities found in transition disks \citep[][]{Espaillat2014}.
However, we note that these emissions are only possible in our model because the dust removal through entrainment with photoevaporative winds is very inefficient, leading to low dust-loss rates, even after considering different entrainment parameters \citep[see \autoref{Sec_Appendix_DustLoss} and][]{Hutchison2016, Hutchison2021, Booth2021}.
If we considered the additional effect of radiation pressure and magneto-thermal wind models to remove solid material \citep[][respectively]{Owen2019, Rodenkirch2022}, we would likely find also a deficit of FIR that is more in line with the predictions for relic disks.\par

Another mechanism that could reduce the disk flux is the conversion of dust particles into planetesimals by streaming instability \citep{Youdin2005}, which tends to limit the optical thickness at the dust ring to values of $\tau_\nu \approx 0.5$ as shown by \citet{Stammler2019}, however this would occur only in the cases where the local dust-to-gas ratio is high with $\epsilon \gtrsim 1$, and would stop as soon as the dust content drops below this threshold. 
Because of the self-regulating nature of planetesimal formation, we do not expect it to significantly reduce the millimeter flux. 
We further note that a recent study  of \citet{Carrera2022} questions the efficiency of the streaming instability as a planetesimal formation scenario in pressure bumps, which would also reinforce our claim.\par

With all these different mechanisms at play it becomes unclear whether we should expect photoevaporating disks to be bright or faint in the FIR.
We propose that the answer depends on the presence or absence of an inner gas disk during the dispersal process, such as the ones sustained by dead zones \citep[][]{Morishima2012, Garate2021}.
In the case that an inner disk is absent, the edge of the photoevaporative cavity (where most of the dust is trapped) should be directly irradiated by the central star, leading to higher photoevaporation rates, and efficient dust clearing by radiation pressure \citep[][]{Picogna2019, Owen2019}. Instead, if an inner disk is present, it could cast a shadow on the outer disk \citep[see][]{Ueda2019} and shield the edge of the photoevaporative cavity from the direct irradiation, slowing its dispersal and reducing the dust loss rates found in \citet[][]{Owen2019}.

Testing this idea however, goes beyond the scope of this work, and requires a proper consideration of the dust distribution of the inner disk in the presence of a dead zone and photoevaporation, the effect of accretion heating which may increase the scale height of the inner disk, and resolving the inner edge of the dead zone where large amounts of dust can be trapped \citep[][]{Ueda2019}.
If the presence of an inner disk during dispersal is correlated with the disk mass \citep[which is likely if the inner disk is sustained by a dead zone, see][]{Turner2007, Delage2022}, then we would expect that more massive disks are more likely to become accreting transition disks that retain a bright dust component, while less massive disks become relic disks that both disperse quickly and lose their dust content due to radiation pressure. 
We expect that only vertically thicker inner disks will be able to block enough stellar irradiation to slow down the dust removal from the outer disk, and it remains to be tested whether or not such inner disk scale heights are achievable.\par

Our preliminary results in \autoref{Sec_Appendix_DustLoss} on this aspect are however inconclusive, as we find that even a perfect entrainment scenario is not enough to completely remove the FIR excess. While dust removal by radiation pressure was shown to reduce the FIR emission by \citet[][see their Figure 11]{Owen2019}, we find that it leads to dust loss rates that are similar to those from our current work for our fiducial parameters. Thus, our model seems unable to create the undetectable relic disks so far.\par

Other avenues to proceed could be to reconsider the relic disk problem in the light of recent results, such as the non-membership of some of the systems considered in the \citet{Hardy2015} study \citep[see][]{Gali2020_GaiaDR2, Luhman2020_GaiaDR2, Michel2021}, and that some of the faint transition disks mentioned in \citet{Owen2012} might be actually inclined disks \citep{vanderMarel2022}, before drawing further conclusions regarding the nature of relic disks.\par

Finally, we note that two recent observations of the disks J16090141 and J16070384 by \citet{vanderMarel2022} could be cataloged as relic disks candidates, since they show most of the expected features, that is, large cavities in the millimeter continuum, low millimeter fluxes, low accretion rates \citep{Alcala2014, Alcala2017}, and low emission in the FIR \citep{Ansdell2018}, though it is still necessary to better characterize the gas depletion of the inner disk in objects before determining if they are compatible with photoevaporative dispersal scenario, and also to consider that these disks are orbiting around low mass stars.

\section{Summary}\label{sec_Summary}

In this work we performed numerical simulations to study the effect that primordial substructures have on the dust evolution and millimeter flux of protoplanetary disks undergoing photoevaporative dispersal.
Our simulations show that the presence of a primordial substructure, specifically in terms of the dust trapping efficiency, determines the flux that the disk displays during its dispersal. Once photoevaporation opens a cavity in the inner regions, further dust loss due to drift is prevented, with dust removal due to wind entrainment having little impact on the evolution of solids.
Therefore, disks that developed early substructures are bright when observed in the millimeter continuum, while disks that were smooth  during its early evolution are faint, with both types of disks maintaining a relatively constant flux as the photoevaporative cavity expands (\autoref{Fig_TotalIntensity}).\par

From our parameter space exploration, we learn that in order to have a bright disk, while undergoing photoevaporative dispersal, the main requirement is that an early substructure with an amplitude of $A_\textrm{gap} \gtrsim 2$ exists (approximately what a Saturn mass planet would cause). Substructures located closer to the star could trap more dust and lead to brighter disks, while higher X-ray luminosities that lead to an earlier dispersal also reduce the amount of dust lost due to diffusion and drift prior to the cavity opening.\par

The millimeter fluxes calculated for the smooth and structured disk models are respectively comparable to those measured in the mm-faint ($F_\textrm{mm} < \SI{30}{mJy}$) and mm-bright ($F_\textrm{mm} > \SI{30}{mJy}$) populations of transition disks, a result that holds for two different opacity prescriptions. 
It is possible then, that at least some observations of transition disks correspond to dispersing disks where early substructures were present (bright disks) or absent (faint disks).\par

While it is not new that the presence of substructures determines the flux of a disk or that photoevaporation leads to the inside-out expansion of a cavity, the combination of both results has a much more distinct consequence:
Bright transition disks with a large cavity are not necessarily caused by massive super-Jupiter mass planets. Instead, these disks can also be created by smaller planets, with masses within the range of Saturn and Jupiter that trap the dusty material at earlier times (without strong constraints on their location), and a photoevaporative dispersal process, where the edge of the expanding cavity drags all the remaining solids in the protoplanetary disk to larger radii.
This scenario could also explain why we have not detected more planets in transition disk observations: they might be both less massive, and located further inside their cavities than previously expected.\par

From the predicted SEDs, we find that our photoevaporating disk models display a bright FIR signal, which could be problematic from the theoretical point of view, since there are no detections of non-accreting transition disks (i.e., relic disks) with  such high FIR fluxes. Studies considering dust removal by radiation pressure \citep[][]{Owen2019}, and long-lived inner disks \citet{Garate2021} that block a fraction of the stellar flux received by the outer disk can alleviate this discrepancy between the current model and observations. Alternatively, surveys that focus on the NIR or MIR emission to detect protoplanetary disks could have missed these transition disks that would display only an FIR component.\par

The simulations presented in this paper, along with the previous work from \citep[][]{Garate2021}, suggest that a comprehensive disk evolution model could explain the observed properties of transition disks, where photoevaporative dispersal is responsible for opening wide cavities, substructures created by planets of moderate mass can trap the dust required to produce the fluxes measured in the millimeter continuum, and dead zones in the inner regions lead long-lived inner disks capable of sustaining the measured accretion rates. The versatility of a combined model opens several pathways to explain the observations of transition disks, and relaxes the constraints imposed on models that rely exclusively on giant planets to explain all the features.\par

To conclude, we highlight the relevance of the synergy between multiple ingredients in disk evolution, in this case photoevaporation and gap-like substructures, which when considered together can explain wider range of the observed features in protoplanetary disks, with fewer constrains.

\begin{acknowledgements}
We would like to thank the anonymous referees for their reports, insights, and suggestions that greatly improved the extent of this paper. We also thank Jochen Stadler for the help with the RADMC3D setup, Nienke van der Marel and James Owen for the insightful discussions, and for sharing a selection of the SEDs and mass loss rates profiles, respectively, that were used for comparison in this paper.
The authors acknowledge funding from the Alexander von Humboldt Foundation in the framework of the Sofja Kovalevskaja Award endowed by the Federal Ministry of Education and Research, from the European Research Council (ERC) under the European Union’s Horizon 2020 research and innovation programme under grant agreement No 714769, by the Deutsche Forschungsgemeinschaft (DFG, German Research Foundation) Ref no. FOR 2634/2 (ER 685/7-1, ER 685/8-2), and by the Deutsche Forschungsgemeinschaft (DFG, German Research Foundation) under Germany's Excellence Strategy – EXC-2094 – 390783311.
\end{acknowledgements}
\bibpunct{(}{)}{;}{a}{}{,} 
\bibliographystyle{aa} 
\bibliography{TheBibliography.bib} 
%
%
%

\begin{appendix}
\section{Dust entrainment prescriptions} \label{Sec_Appendix_DustLoss}

\subsection{Entrainment parameters}

\begin{figure}
\centering
\includegraphics[width=90mm]{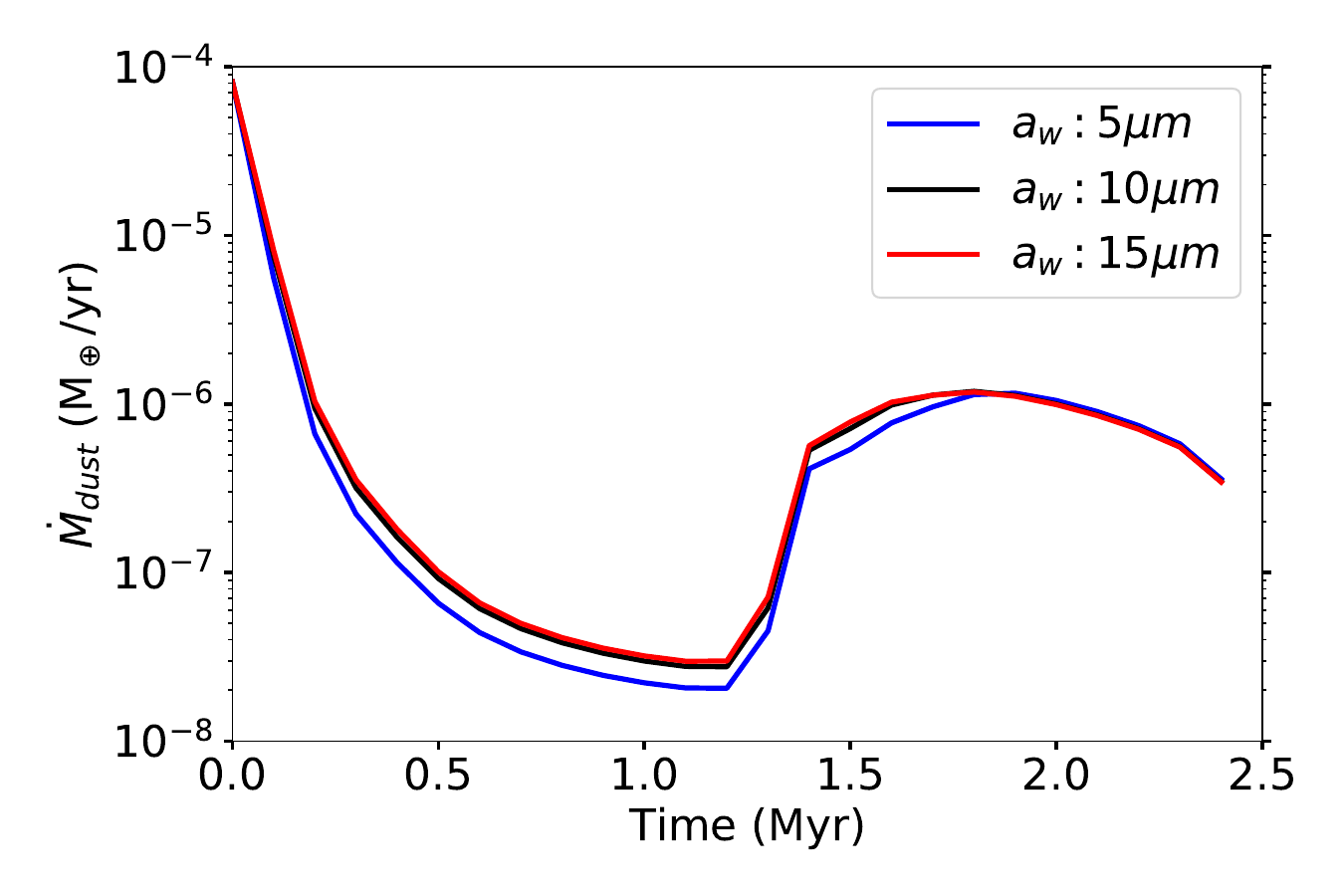}
\includegraphics[width=90mm]{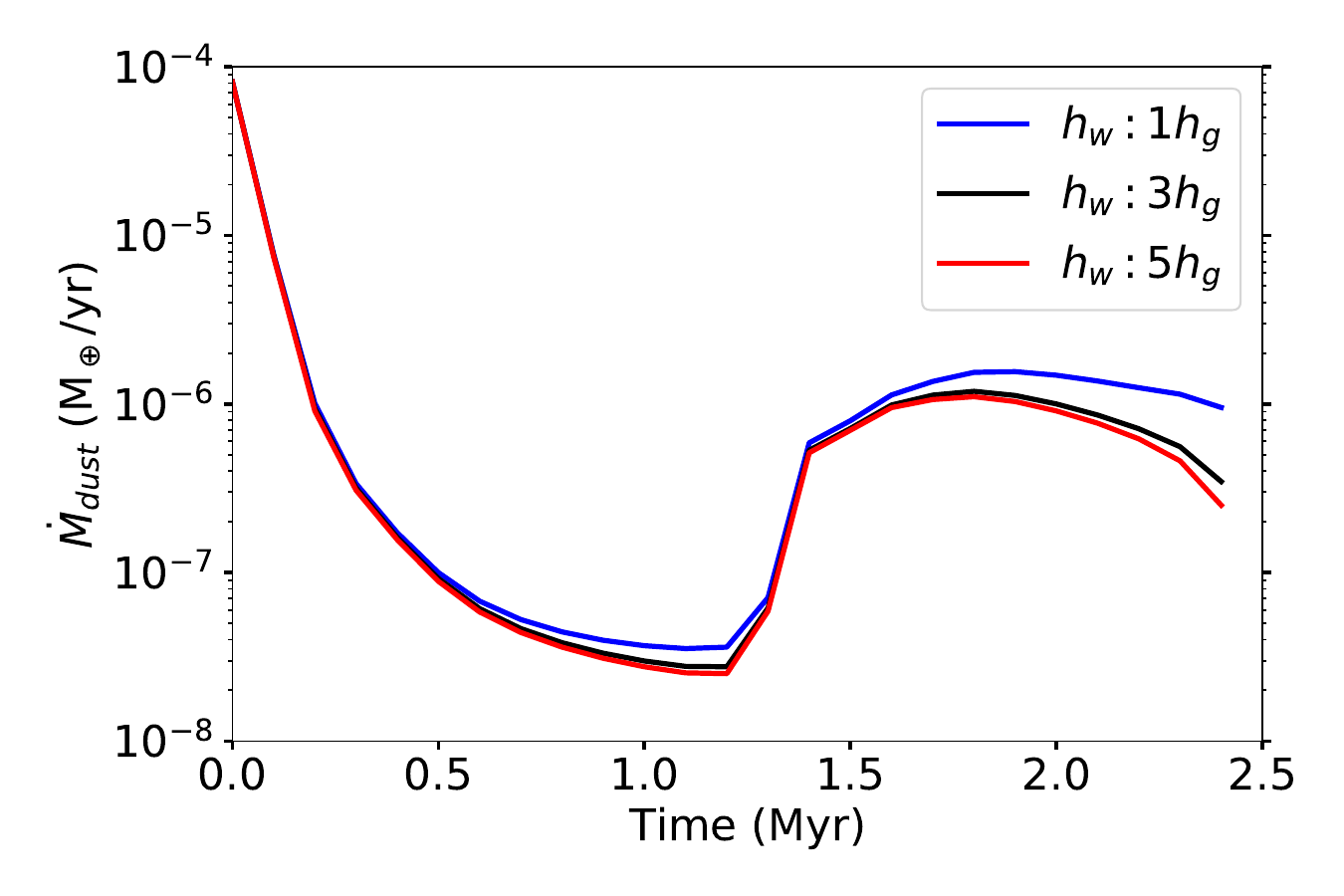}
 \caption{Evolution of the dust loss rate for the fiducial smooth disk model, using different entrainment parameters $a_\textrm{w}$ (\textit{top}) and $h_\textrm{w}$ (\textit{bottom}).
 }
 \label{Fig_Appendix_EntrainmentSmooth}
\end{figure}

For this work, we used a prescription of dust entrainment that removes dust grains smaller than a certain size $a_\textrm{w}$, located above a certain scale height $h_\textrm{w}$.
The values used and the mass loss rates obtained in this work are consistent with the values found in the studies of dust entrainment by \citet{Franz2020} and \citet{Franz2022a}. However, our implementation is still a simplification of the of dust dynamics in the wind launching region, as the maximum entrainment size and scale height of the launching wind region change with radius.
To address this limitation, we perform additional simulations with varying the two parameters $a_\textrm{w}$ and $h_\textrm{w}$, and compare the evolution of the dust loss rate over time. The remaining parameters are the same as in the fiducial smooth disk simulation.\par

\autoref{Fig_Appendix_EntrainmentSmooth} shows that neither the variations in the entrainment scale height, nor in the maximum entrainment size have a major impact on the dust loss rate evolution, which is expected, since most of the mass loss contribution comes from the smaller grain sizes.
Only when the entrainment scale height is $h_\textrm{w} = 1 h_\textrm{g}$ we observe a constant dust loss rate toward the end of the disk lifetime, however we note that this value is highly unlikely, since \citet{Franz2020} and \citet{Rodenkirch2022} show that the wind launching region is located between $3.5 h_\textrm{g}$ and $5.5 h_\textrm{g}$.
Based on this parameter study, we do not expect for the exact values of the entrainment prescription to influence our results.


\subsection{Comparison with Booth and Clarke (2021) entrainment.}

\begin{figure}
\centering
\includegraphics[width=90mm]{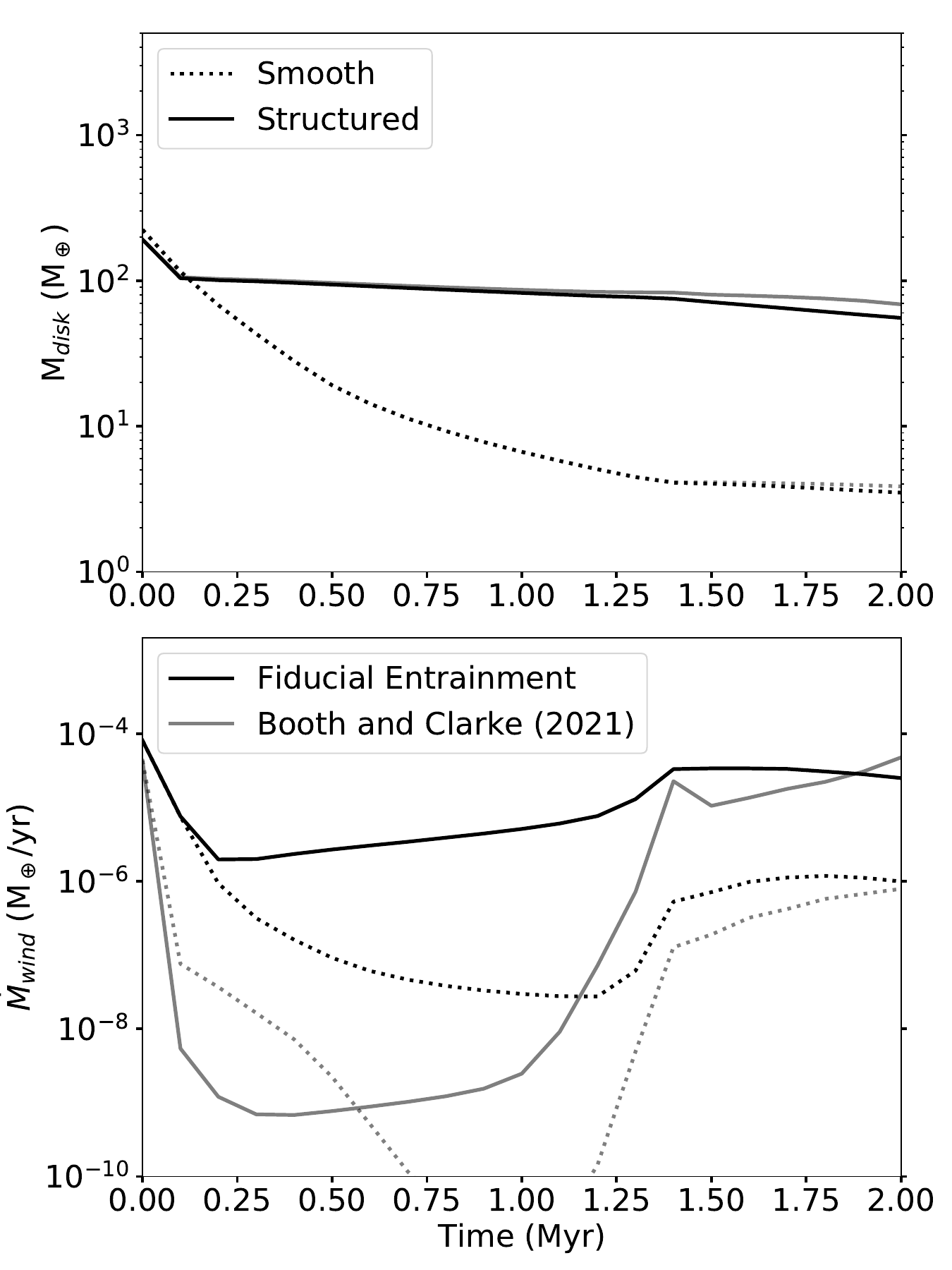}
 \caption{Comparison between the evolution of the dust mass (\textit{top}) and dust loss rate (\textit{bottom}), for the fiducial disk models (\textit{black}) against the entrainment prescription by \citet[][\textit{gray}]{Booth2021}.
 }
 \label{Fig_Appendix_EntrainmentBooth}
\end{figure}

\begin{figure}
\centering
\includegraphics[width=90mm]{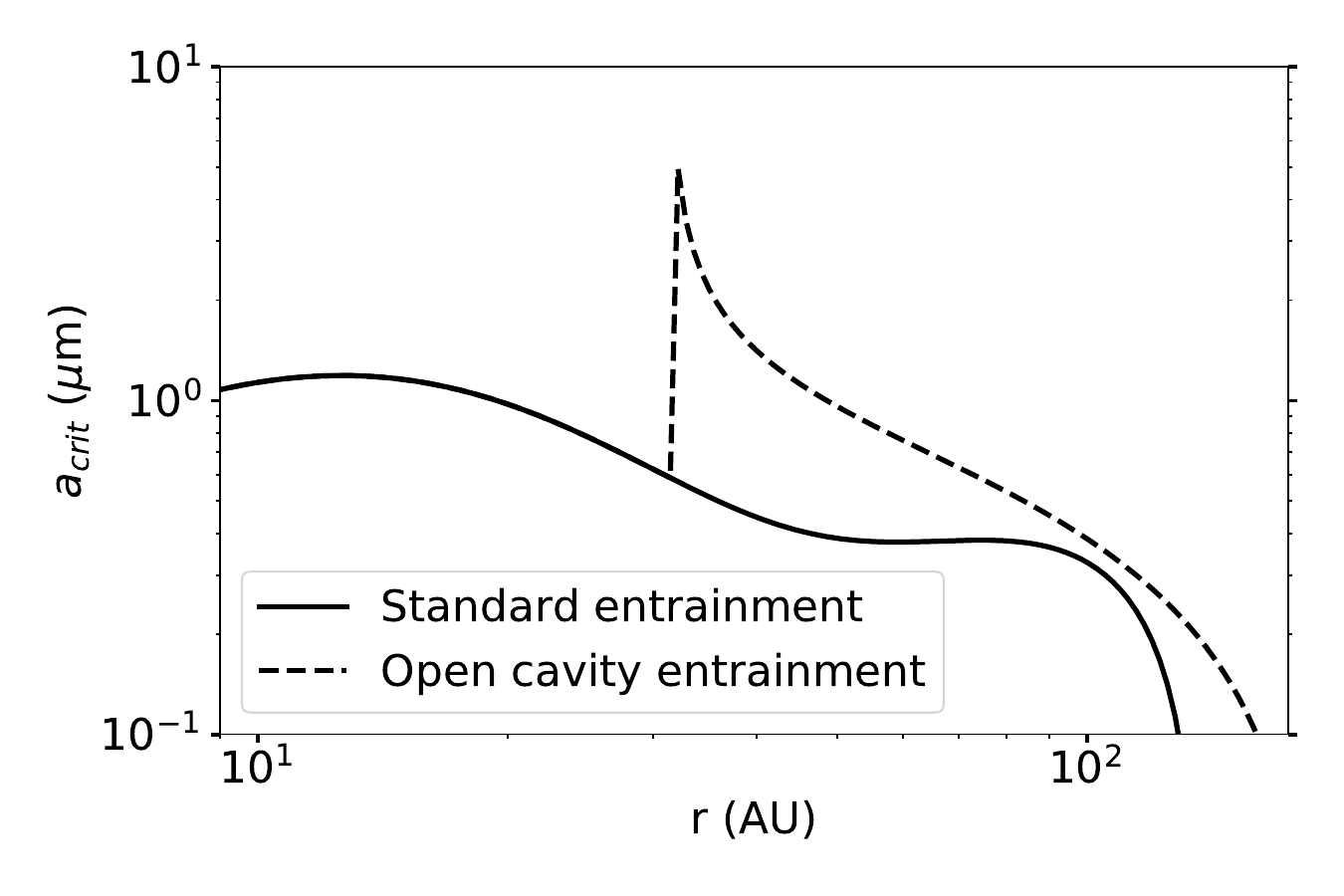}
 \caption{Dust entertainment size given by the \citet{Booth2021} prescription, for the standard photoevaporative mass loss rate (solid) and the open cavity mass loss rates (dashed), for the fiducial X-ray luminosity of $L_x = \SI{e30}{erg\, s^{-1}}$.
 }
 \label{Fig_Appendix_EntrainmentBooth_size}
\end{figure}

The study of \citet{Booth2021} shows that solids can be lifted to the disk surface due to an efficient advection triggered by the photoevaporative mass loss, which may make the dust grains more easily entrained with the photoevaporative wind.
To compare how their prescription, which defines the effective size for dust grains are entrained would affect our model, we run two additional simulations based on our fiducial setup (with and without early dust traps), implementing their equation 25:
\begin{multline}
    a_\textrm{crit} = 0.63 \left(\frac{\dot{\Sigma_\textrm{w}}}{\SI{e-12}{g\, cm^{-2} s^{-1}}} \right) \left(\frac{r}{\SI{10}{AU}} \right)^{3/2} \\
    \times \left(\frac{z_\textrm{IF}}{4 H_\textrm{IF}} \right)^{-1} \left(\frac{M_*}{M_\odot} \right)^{-1/2} \left(\frac{\rho_s}{\SI{1}{g cm^{-3}}} \right)^{-1} \mu\textrm{m},
\end{multline}
assuming a launching height of $z_\textrm{IF} = 4 H_\textrm{IF}$. This prescription then replaces our dust-to-gas loss ratio expression described in \autoref{eq_lossratio}.
\par

The evolution of the mass and mass loss rate using the \citet{Booth2021} prescription is shown in \autoref{Fig_Appendix_EntrainmentBooth}, where we do not notice almost any differences from our fiducial models in terms of total mass of solids, and observe that the mass loss rates derived from their more accurate entrainment prescription are even lower than those considered in this work, making them negligible for the global disk evolution.
\par

The lower mass loss rates can be explained by looking at the entrainment sizes derived from the \citet{Booth2021} prescription, which are typically lower than those of our fiducial model, as shown in \autoref{Fig_Appendix_EntrainmentBooth_size}.

%
\subsection{Maximum dust entrainment}

\begin{figure}
\centering
\includegraphics[width=90mm]{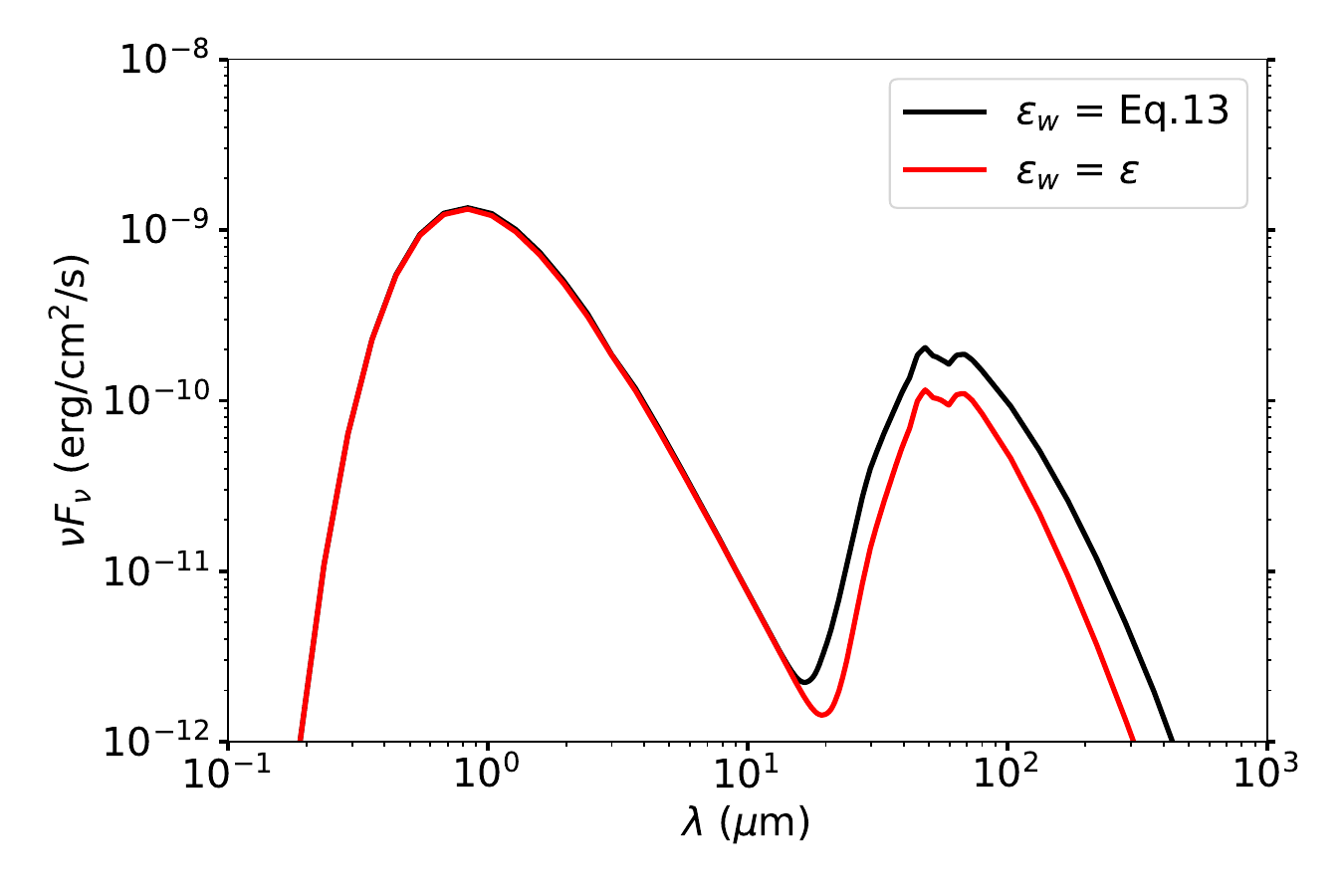}
 \caption{SED for the smooth disk model with two different prescriptions for the dust entrainment $\epsilon_w$.
 The snapshots are taken when the cavity size is approximately $r_\textrm{cavity} = \SI{100}{AU}$, assuming a distance of $d = \SI{140}{pc}$, and no inclination.
 }
 \label{Fig_Appendix_SED_Entrainment}
\end{figure}

\begin{figure}
\centering
\includegraphics[width=90mm]{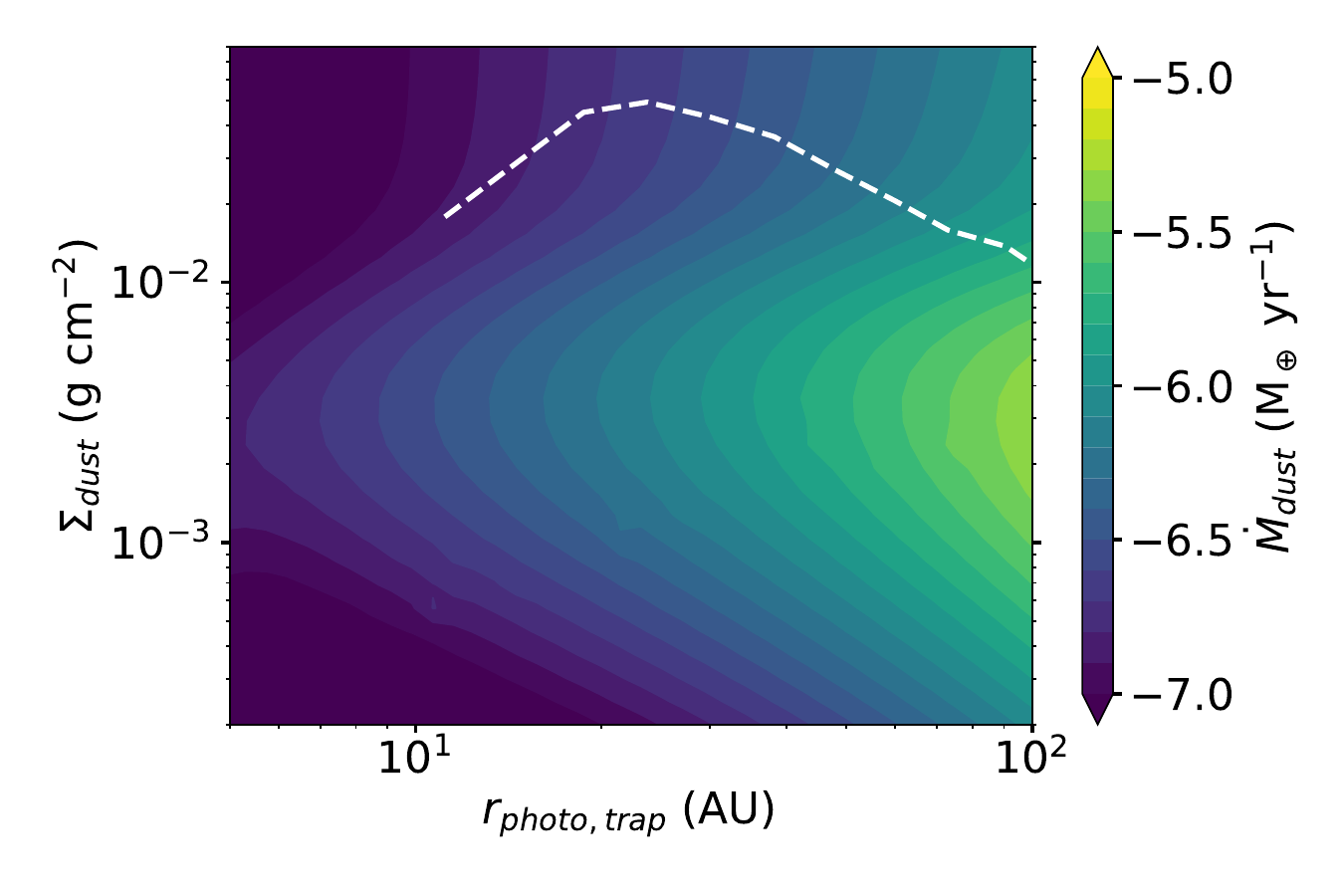}
\includegraphics[width=90mm]{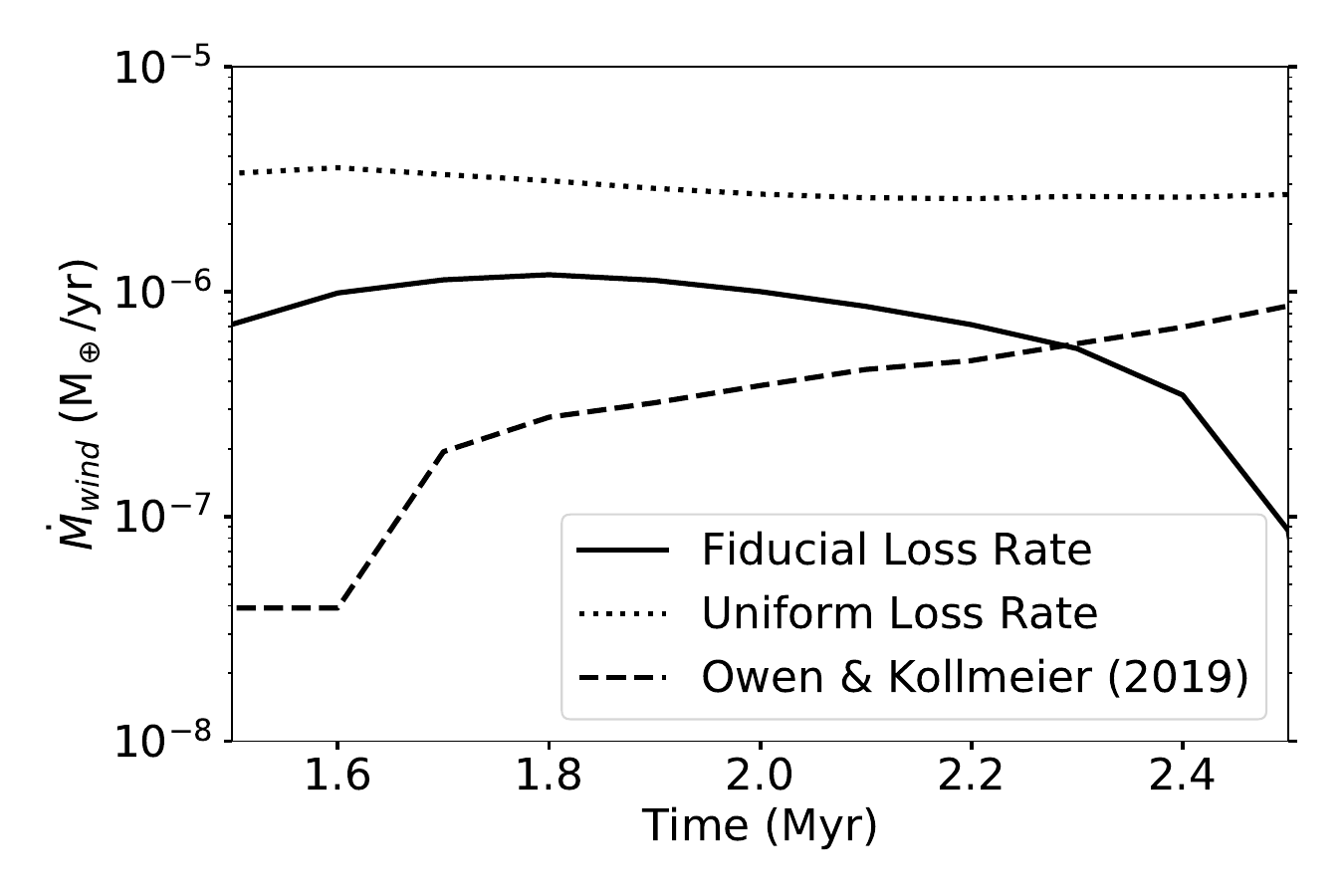}
 \caption{
 \textit{Top:} Mass loss rates from \citet[][Figure 6]{Owen2019} due to radiation pressure, with the photoevaporative trap location and dust surface density evolution of our fiducial model plotted on top (dashed line).
 \textit{Bottom:} Mass loss rates for the fiducial smooth disk model (solid), the disk model with uniform entrainment ($\epsilon_w = \epsilon$, dotted), and the estimated loss rates from the \citet{Owen2019} radiation pressure (dashed), given our fiducial model properties.
 }
 \label{Fig_Appendix_Owen_Entrainment}
\end{figure}

A limit test of dust entrainment in photoevaporative winds would be the case in which the dust loss rate is always proportional to the local (vertically integrated) dust-to-gas ratio, with $\dot{\Sigma}_\textrm{d} = \epsilon \dot{\Sigma}_\textrm{g}$, which ignores the effects of growth and settling.
This prescription would lead to the highest possible dust loss rate from entrainment alone, and to the faintest emission from the dust in IR that can be obtained from this model.\par

\autoref{Fig_Appendix_SED_Entrainment} shows a comparison between the resulting SED from our fiducial smooth disk simulation when considering the standard entrainment prescription (\autoref{eq_lossratio}), against the limit case when $\epsilon_\textrm{w} = \epsilon$. 
We see that the emission peaks between $\SI{40}{\mu m}$ and $\SI{70}{\mu m}$, with a value of $F_\textrm{IR} \approx \SI{e10}{erg\, cm^{-2}\, s^{-1}}$ when the $\epsilon_\textrm{w} = \epsilon$ prescription is considered, which is approximately half of the emission from the standard entrainment prescription (\autoref{eq_lossratio}).\par

Nevertheless, the IR emission is still higher than the median of class II objects from \citet[][]{Ribas2017}, and therefore should also be detectable with current instrumentation, despite the enhanced dust entrainment and loss rates. In other words, this object could not represent one of the yet un-detected relic disks \citep[][]{Owen2016_review}.\par
In order to compare our mass loss rates with those found by \citet{Owen2019}, which include dust removal by radiation pressure, we map the location of the photoevaporative dust trap and dust surface density into the mass loss grid shown in \autoref{Fig_Appendix_Owen_Entrainment} (top panel), and extract the corresponding mass loss rate of solids.
We find the expected dust loss rates by radiation pressure are comparable or lower than those found in our fiducial model, and we do not expect them to remove a significant amount of dust for our current setup. 
Higher dust loss rates could be reached if the dust surface densities were lower, as the mass loss grid reaches its maximum at values of $\Sigma_\textrm{dust} \sim \SI{3e-3}{g\, cm^{-2}}$, however it seems unlikely that the additional contribution of radiation pressure alone can remove enough dust to completely dim the FIR emission, given our radiative transfer calculations of the SED emission shown in \autoref{Fig_Appendix_SED_Entrainment}.

\end{appendix}

\end{document}